\documentclass[12pt]{article}

\pdfoutput=1
\usepackage[latin9]{inputenc}
\usepackage[top=80pt,bottom=85pt,left=55pt,right=55pt]{geometry}
\usepackage{amssymb,amsmath}
\usepackage{xypic,subfigure}
\usepackage{graphicx,color}
\usepackage{float}
\usepackage{cite}
\usepackage[debug,pageanchor=false]{hyperref}
\definecolor{link}{rgb}{.8,.15,.1}
\hypersetup{colorlinks=true,linkcolor=link,citecolor=link,urlcolor=link,linktocpage}
\usepackage{hyperref}
\makeatletter
\@addtoreset{equation}{section}
\makeatother

\renewcommand{\theequation}{\thesection.\arabic{equation}}

\newcommand{\beq}{\begin{equation}}
\newcommand{\eeq}{\end{equation}}
\newcommand{\bea}{\begin{eqnarray}}
\newcommand{\eea}{\end{eqnarray}}
\newcommand{\nn}{\nonumber}

\newcommand{\eq}{\begin{equation}}
\newcommand{\feq}{\end{equation}}
\newcommand{\eqn}{\begin{eqnarray}}
\newcommand{\feqn}{\end{eqnarray}}

\def\e{\mathrm{e}}

\newcommand{\be}{\begin{equation}}
\newcommand{\ee}{\end{equation}}
\newcommand{\dd}{d}
\newcommand{\me}{e}

\newcommand{\vol}{\mathrm{vol}}

\begin{document}
\begin{titlepage}

\begin{center}

\vskip .5in 
\noindent

{\Large \bf{A plethora of Type IIA embeddings for $d=5$ minimal supergravity}}

\bigskip\medskip

Christopher Couzens$^{a}$\footnote{cacouzens@khu.ac.kr},  Niall T. Macpherson$^{b,c}$\footnote{ntmacpher@gmail.com}, 	Achilleas Passias$^{d}$\footnote{achilleas.passias@lpthe.jussieu.fr} \\

\bigskip\medskip
{\small 

$a$: Department of Physics and Research Institute of Basic Science,\\
  Kyung Hee University, Seoul 02447, Republic of Korea\\
  
\vskip 3mm
$b$: Departamento de F\'isica de Part\'iculas\\
 Universidade de Santiago de Compostela\\
	and\\
	$c$: Instituto Galego de F\'isica de Altas Enerx\'ias (IGFAE)\\
	R\'ua de Xoaqu\'in D\'iaz de R\'abago s/n\\
	E-15782 Santiago de Compostela, Spain

\vskip 3mm
$d$: Sorbonne Universit\'{e}, UPMC Paris 06, UMR 7589, LPTHE, \\ 75005 Paris, France  \\
}
\bigskip\medskip

\vskip 1cm 

     	{\bf Abstract }
     	\end{center}
     	\noindent
	
We construct multiple embeddings of all solutions of $d=5$ minimal (un)gauged supergravity into massive Type IIA supergravity. The internal spaces and warpings of such embeddings are the same as those of the $\mathcal{N}=1$ supersymmetric (Mink$_5$) AdS$_5$ vacua, with the slight modification that the U(1) R-symmetry direction becomes fibered over the external space by the $d=5$ gauge field. In addition the fluxes are appropriately modified. There are many distinct types of the aforementioned internal spaces and as such many different embeddings of the $d=5$ supergravity. As examples of our setup we provide new solutions dual to six-dimensional, $\mathcal{N}=(1,0)$ SCFTs compactified on the product of a constant curvature Riemann surface and a spindle. We also provide a multitude of massive Type IIA embeddings for rotating, asymptotically AdS$_5$ black hole solutions.

\noindent

\vfill
\eject

\end{titlepage}

\setcounter{footnote}{0}

\tableofcontents

\setcounter{footnote}{0}
\renewcommand{\theequation}{{\rm\thesection.\arabic{equation}}}

\section{Introduction}

Supersymmetric solutions of supergravity theories play a key role in the study of string theory. Supersymmetric compactifications provide a setting for obtaining realistic models of particle physics, while a microscopic derivation of the black hole entropy in string theory is best understood for supersymmetric black holes. Supersymmetric solutions have also supported the development of the gauge/gravity duality.

Supersymmetry is, in part, technically a simplifying assumption in the construction of solutions. Still, especially in the absence of a high degree of supersymmetry or other symmetries, the construction of solutions of ten- or eleven-dimensional supergravity theories is a challenging task which calls for advanced mathematical tools, especially in the field of geometry. 

Complementary to the construction of solutions directly in ten or eleven dimensions, is the uplift of solutions of lower-dimensional supergravity theories. The latter is feasible due to the existence of consistent truncations of the infinite Kaluza--Klein tower of higher-dimensional compactifications to a finite set of modes, so that a solution of the lower-dimensional equations of motion is also a solution of the ten- or eleven-dimensional ones. Examples include consistent truncations on spheres down to maximal gauged supergravities \cite{Nastase:1999cb,Nastase:1999kf,Cvetic:2000nc,Baguet:2015sma,Guarino:2015vca,deWit:1986oxb,Varela:2015ywx}, Sasaki--Einstein manifolds \cite{Gauntlett:2009zw, Cassani:2010uw, Gauntlett:2010vu, Liu:2010sa}, weak-$G_2$ holonomy manifolds \cite{Gauntlett:2009zw} and tri-Sasakian manifolds \cite{Cassani:2011fu}, SU(2)-structure \cite{Triendl:2015rta, Larios:2019lxq} and SU(3)-structure \cite{Cassani:2012pj} manifolds, as well as spaces including brane singularities \cite{Passias:2015gya}. Recently, a framework based on exceptional generalised geometry and exceptional field theory has emerged, that allows for a systematic treatment of consistent truncations \cite{Malek:2018zcz,Malek:2019ucd,Malek:2020jsa,Josse:2021put,Galli:2022idq}. Despite these successes the exceptional field theory framework for consistent truncations has only been fully worked out for reductions to maximal and half-maximal gauged supergravities.\footnote{That being said, in \cite{Cassani:2019vcl} (which generalises the earlier half-maximal work of \cite{Malek:2017njj}) there is a generic prescription to obtain a consistent truncation preserving any amount of supersymmetry.}

In the present work we construct a universal consistent truncation of massive Type IIA supergravity on a five-dimensional Riemannian manifold $M_5$, to minimal (un)gauged supergravity in five dimensions. The manifold $M_5$ can be any of the class of manifolds that constitute the internal space of five-dimensional, $\mathcal{N}=1$ supersymmetric Minkowski (Mink$_5$) or anti-de Sitter (AdS$_5$) solutions of massive Type IIA supergravity \cite{Apruzzi:2015zna}.\footnote{The corresponding truncation of Type IIB supergravity was carried out in \cite{Gauntlett:2007ma}, (see also \cite{Gauntlett:2006ai}). Our work includes the embeddings of \cite{Cheung:2022wpg}.} We apply the technical methodology of that work, the bi-spinor formalism in conjunction with $G$-structures, to the construction of the consistent truncation Ansatz.\footnote{See also \cite{Rosa:2013jja}.}

Five-dimensional minimal supergravity is a rich theory \cite{Gauntlett:2003fk}, and our work paves the way for the uplift of many interesting solutions that reside in it (e.g. \cite{Gutowski:2004ez}), in a multitude of ways, and their subsequent study in Type IIA supergravity. We consider the uplift of two classes of solutions as examples of our consistent truncation. The first includes the solution of \cite{Gutowski:2004ez} describing a black hole with two independent angular momenta and a single magnetic charge. The second is the near-horizon of a black string which has a spindle horizon, first studied in \cite{Ferrero:2020laf} where it was uplifted to Type IIB supergravity on a Sasaki--Einstein manifold. Later work has generalised the spindle solutions to different dimensions $\geq4$ and different embeddings in string/M-theory, \cite{Ferrero:2020twa,Hosseini:2021fge,Boido:2021szx,Faedo:2021kur,Ferrero:2021wvk,Cassani:2021dwa,Ferrero:2021ovq,Couzens:2021rlk,Ferrero:2021etw,Couzens:2021cpk,Faedo:2021nub,Giri:2021xta,Cheung:2022ilc,Suh:2022olh,Arav:2022lzo,Couzens:2022yiv}. 
~\\
The rest of the paper is organised as follows:

In section \ref{eq:vaccua} we lay down the groundwork to embed  $d=5$ minimal (un)gauged supergravity into massive Type IIA supergravity. We discuss the (Mink$_5$) AdS$_5$ vacua of massive Type IIA supergravity with generalised structures in section \ref{eq:generalisedstructures}, which reviews and slightly generalises (allowing for Mink$_5$) the results of \cite{Apruzzi:2015zna}. An important part of this section for our later generalisation is appreciating that AdS$_5$ vacua support both null and time-like Killing vectors (in the sense of \cite{Tomasiello:2011eb}), with the latter yielding information pertinent to our ultimate aim more readily.  We review the known class of AdS$_5$ vacua in section \ref{sec:onAdS5vaccua}, writing them in a convenient form for our later purposes. In section \ref{sec:onmink5vaccua} we derive a new class of Mink$_5$ vacua, relevant for the ungauged limit of the $d=5$ supergravity.

In section \ref{eq:theembedding} we derive an embedding of  $d=5$ minimal (un)gauged supergravity into massive Type IIA supergravity under the assumption that the ten-dimensional solution decomposes as a warped product with the five-dimensional U(1) gauge field appearing as a connection in the ten-dimensional metic. We further assume that the ten-dimensional  bosonic fields depend on $d=5$ minimal (un)gauged supergravity only through its bosonic fields --- so we can obtain an embedding that does not depend on external supersymmetry. To derive the embedding we make use of the same language of generalised structures used to derive the (Mink$_5$) AdS$_5$ vacua, a major benefit being that there is no need to make an ansatz for the flux, which is uniquely fixed by our previous assumptions. We show that the internal space of the ten-dimensional solutions is a mild generalisation of that of (Mink$_5$) AdS$_5$ vacua and provide simple replacement rules to map a ten-dimensional vacuum solution to an embedding of a generic solution of $d=5$ minimal (un)gauged supergravity. Given a solution to $d=5$ supergravity, there are as many embeddings as there are ten-dimensional vacua, i.e.\ many. This section is supplemented by appendix \ref{sec:suff} where we prove that for any of these embeddings ten-dimensional supersymmetry is preserved whenever five-dimensional supersymmetry holds and that one has a solution to the ten-dimensional equations of motion regardless.

In section \ref{sec:examples} we uplift two classes of solutions to massive Type IIA supergravity. The seed AdS$_5$ solutions were constructed in \cite{Bah:2017wxp} and consist of a constant curvature Riemann surface present in the internal space as well as an O8--D8 stack, D6-brane and D4-brane sources, localized and partially localized. The solutions are characterised by a cubic polynomial with different global completions depending on the choice of four parameters. The solutions have the natural interpretation of being the holographic duals of the four-dimensional superconformal field theories (SCFTs) arising in the IR limit of placing a six-dimensional, $\mathcal{N}=(1,0)$ theory on the Riemann surface. Using the seed solutions we show how to uplift two classes of solutions of $d=5$ minimal gauged supergravity to the massive Type IIA one. The first class of solutions is the Gutowski--Reall black hole solutions with equal angular momenta parameters \cite{Gutowski:2004ez}; one could also use our formulae to uplift the CCLP solution \cite{Chong:2005hr} which has two independent angular momenta. The second class are the AdS$_3\times \mathbb{WCP}^{1}_{n_{\pm}}$ spindle solutions \cite{Ferrero:2020laf}, which also include in a particular limit the AdS$_3\times \Sigma_{g>1}$ solutions. 

The work is supplemented by technical appendices referred to in the main text.

\section{Generalised structures and vacua}\label{eq:vaccua}
Before proceeding with the embedding of $d=5$ minimal (un)gauged supergravity into massive Type IIA supergravity, it will be useful to review some features of the $\mathcal{N}=1$ supersymmetric AdS$_5$ vacua of the latter. These were originally classified in \cite{Apruzzi:2015zna}, with their local form significantly refined in \cite{Bah:2017wxp}. Something that will be particularly useful going forward is how these vacua arise from the necessary and sufficient conditions for supersymmetry phrased in terms of generalised structures in $d=10$ \cite{Tomasiello:2011eb}.

\subsection{Bi-spinor equations}\label{eq:generalisedstructures}
The fundamental objects appearing in the classification of \cite{Tomasiello:2011eb} are bi-linears of the two Majorana--Weyl supersymmetry parameters of Type II supergravity $\epsilon_{1,2}$, namely
\begin{align}
K^{(10)}&\equiv\frac{1}{64}(\overline{\epsilon}_{1}\Gamma_M \epsilon_{1}+\overline{\epsilon}_{2}\Gamma_M \epsilon_{2})dx^M,~~~~\tilde{K}^{(10)}\equiv\frac{1}{64}(\overline{\epsilon}_{1}\Gamma_M \epsilon_{1}-\overline{\epsilon}_{2}\Gamma_M \epsilon_{2})dx^M,~~~~
\Psi^{(10)}\equiv \epsilon_1\otimes \overline{\epsilon}_2.
\end{align}
Necessary conditions for supersymmetry are given in terms of these and the dilaton, NSNS 3-form and RR polyform, respectively $(\Phi,H,F)$, 
\begin{subequations}
\begin{align}
d_H(e^{-\Phi}\Psi^{(10)})&=- (\iota_{K^{(10)}}+\tilde{K}^{(10)}\wedge)F,\label{eq:10dsusy1}\\[2mm]
\nabla_{(M} K^{(10)}_{N)}&=0,~~~~ d \tilde{K}^{(10)}= \iota_{K^{(10)}}H\label{eq:10dsusy2},
\end{align}
\end{subequations}
where $d_H\equiv d-H\wedge$. These conditions imply that 
\beq\label{eq:decendent}
{\cal L}_{K^{(10)}}\Psi^{(10)}={\cal L}_{K^{(10)}}\Phi=0,
\eeq
and further, when the  Bianchi identities for the fluxes are assumed, namely
\beq
dH=0,~~~~ d_H F=0,
\eeq
that ${\cal L}_{K^{(10)}}H={\cal L}_{K^{(10)}}F=0$. Thus,  $(K^{(10)})^M\partial_M$ is an isometry of any supersymmetric solution, under which $\epsilon_{1,2}$ are singlets. The conditions \eqref{eq:10dsusy1}-\eqref{eq:10dsusy2} are not by themselves sufficient for supersymmetry generically; for that one must also solve some so called pairing constraints --- however, for the cases we are interested in they are actually implied so we shall not quote them here.\\
~~\\
An AdS$_5$ vacuum solution of massive Type IIA supergravity must have bosonic fields decomposing as
\beq\label{eq:AdS5vacume}
ds^2_{10}=e^{2A} ds^2(\text{AdS}_5)+ ds^2(\text{M}_5),~~~~~F= f_++ e^{5A}\text{vol}(\text{AdS}_5)\wedge \star\lambda(f_+),
\eeq
where $(e^{2A},f_+)$ have support on M$_5$ only and likewise for the NSNS 3-form and dilaton, while we assume AdS$_5$ has inverse radius $m$.\footnote{Note: By definition $F= \sum_{k=0}^5 F_{2k}$, $f_+=F_0+ f_2+ f_4$ is the magnetic part of this RR polyform, while $\lambda(C_k)=  (-1)^{\lfloor\frac{k}{2}\rfloor} C_k$, for a $k$-form $C_k$.} When such vacua are supersymmetric they can be extracted from \eqref{eq:10dsusy1}-\eqref{eq:10dsusy2} by decomposing the $d=10$ Killing spinors as 
\beq 
\epsilon_1= \frac{1}{\sqrt{2}}\left(\begin{array}{c}1\\
 i \end{array}\right)\otimes \zeta \otimes \chi_1+\text{m.c.},~~~~\epsilon_2= \frac{1}{\sqrt{2}}\left(\begin{array}{c}1\\ -i\end{array}\right)\otimes \zeta\otimes \chi_2+\text{m.c.},\label{eq:spinoransatz}
\eeq
where $\chi_{1,2}$ are Dirac spinors on the internal space,  $\zeta$ are Killing spinors on AdS$_5$ obeying
\beq\label{eq:KSE}
\nabla_{\mu}\zeta= \frac{m}{2} \gamma_{\mu}\zeta,
\eeq
and here and elsewhere m.c.\ stands for Majorana conjugate. Our conventions for gamma matrices can be found in appendix \ref{sec:appendix}. Defining  bi-spinors on AdS$_5$ and M$_5$ as
\begin{align}
\phi^1&= \zeta\otimes \overline{\zeta},~~~~~ \phi^2= \zeta\otimes \overline{\zeta^c},\nn\\[2mm]
\psi^1&=\chi_1\otimes \chi_2^{\dag},~~~~\psi^2= \chi_1\otimes \chi_2^{c\dag},
\end{align}
where $\phi^2$ only has non-trivial 2- and 3-form contributions, 
one finds that the $d=10$ bi-linears decompose as
\begin{align}
K^{(10)}&=\frac{1}{16}(q_+ k- f \xi),~~~~\tilde{K}^{(10)}=\frac{1}{16}(q_- k-f \tilde{\xi}),~~~q_{\pm}= \frac{e^{A}}{2}(|\chi_1|^2\pm |\chi_2|^2),\nn\\[2mm]
\Psi^{(10)}&=\left(i \phi^1_0\text{Im}\psi^1_++e^{5A}\phi^1_5\wedge \text{Im}\psi^1_-\right)+ e^{A}\phi^1_1\wedge \text{Im}\psi^1_-+e^{2A}\left(\phi^1_2\wedge \text{Re}\psi_+^1+\text{Im}(\phi^2_2\wedge \psi^2_+)\right)\nn\\[2mm]
&-e^{3A}\left(i\phi^1_3\wedge \text{Re}\psi^1_-+\text{Re}(\phi^2_3\wedge \psi^2_-)\right)+ie^{4A}\phi^1_4\wedge \text{Im}\psi^1_+\label{eq:10dbilinears},
\end{align}
where we introduced the following real function $f$ and real 1-forms ($k$, $\xi$, $\tilde{\xi}$):
\beq
f\equiv -i 4\phi^1_0,~~~~k\equiv 4 \phi^1_1,~~~~\xi\equiv \frac{1}{2}(\chi_1^{\dag}\gamma_{\underline{a}} \chi_1-\chi_2^{\dag}\gamma_{\underline{a}} \chi_2)\e^{\underline{a}},~~~~\tilde{\xi}\equiv \frac{1}{2}(\chi_1^{\dag}\gamma_{\underline{a}} \chi_1+\chi_2^{\dag}\gamma_{\underline{a}} \chi_2)\e^{\underline{a}}\label{eq:somedefs}.
\eeq  
It is a simple application of Fierz identities to establish that \eqref{eq:KSE} implies the following equations\footnote{Here (deg) indicates that the form degree appears here, i.e.\ $(\text{deg})C_k=k C_k$.}
\beq
d\phi^1_-=m (\text{deg})\phi^1_+,~~~~d\phi^1_+=0,~~~~d\phi^2_2=3m \phi^2_3,~~~~d\phi^2_3=0,~~~~\nabla_{(\nu}k_{\nu)}=0,\label{eq:5ddiffbi}
\eeq
so that in particular  $f$ is constant and $k^{\mu}\partial_{\mu}$ is a Killing vector. One can show in general\footnote{I.e. this follows from the generic properties of a Lorentzian bi-linear in $d=5$.} that $\iota_k k=-f^2$ and it follows from \eqref{eq:5ddiffbi} (given identities in appendix \ref{eq:deq5lordetails})  that
\beq\label{eq:trans}
{\cal L}_k\phi^1=0,~~~~{\cal L}_k \phi^2= 3i m f \phi^2,
\eeq
so the nature of $k^{\mu}\partial_{\mu}$, null/time-like, singlet/charged is intimately related to the value of $f$. There are of course two types of supercharges that AdS$_5$ preserves: Poincar\'e supercharges $\zeta_P$ and conformal supercharges $\zeta_C$.\footnote{If one parametrises AdS$_5$ as $ds^2({\rm AdS}_5) = e^{2m r} (dx^{\alpha})^2+dr^2$ for $\alpha=0,...,3$, then in the obvious frame this suggests, these are $\zeta_P= e^{\frac{1}{2}m r} \zeta^0_+$ and $\zeta_C=(e^{-\frac{1}{2}m r}+ m e^{\frac{1}{2}m r} x^{\underline{\alpha}} \gamma_{{\underline{\alpha}}})\zeta^0_-$ where $\zeta^0_{\pm}$ are constant spinors obeying $\gamma_{\underline{r}}\zeta^0_{\pm}=\pm \zeta^0_{\pm}$.} We can choose to align $\zeta$ along any (non-zero) linear combination of these without changing anything physical about the AdS$_5$ vacua,  however taking without loss of generality
\beq
\zeta= \zeta_P+i \zeta_C~~~~\Rightarrow~~~~~ f= 2\text{Re}(\overline{\zeta}_P\zeta_C),
\eeq
so $f$ is only non-zero if we align $\zeta$ along both such charges and we can in fact extract information more easily from \eqref{eq:10dbilinears} by making this choice. To see this one can consider for instance \eqref{eq:10dsusy2}:  plugging \eqref{eq:10dbilinears} into this one finds it requires
\beq\label{eq:zeroforms}
e^{-2A}q_+=c,~~~~m q_-=0,~~~~dq_-=0,~~~~ f \nabla_{(a}\xi_{b)}=0,~~~~f(d\tilde\xi-\iota_{\xi}H)=0,
\eeq
where $c>0$ is a constant, so \eqref{eq:10dsusy2} imply that $\xi^a\partial_a$ is a Killing vector and fixes the part of $H$ parallel to it but only when $f\neq0$. Of course as nothing physical should depend on how $\zeta$ is parameterised, and hence the value of $f$, $\xi^a\partial_a$ should always be Killing --- indeed \cite{Apruzzi:2015zna}, which implicitly assumes $f=0$, show this explicitly, albeit with a less direct computation. Likewise given \eqref{eq:trans} the first of \eqref{eq:decendent} imposes
\beq
f( {\cal L }_{\xi}\psi^1)=f( {\cal L}_{\xi}\psi^2- 3i m c \psi^2)=0,
\eeq
making clear that $\xi^a\partial_a$ is the U(1) R-symmetry one expects an $\mathcal{N}=1$ supersymmetric AdS$_5$ solution to support.
Finally, plugging \eqref{eq:10dbilinears} and the second of \eqref{eq:AdS5vacume} into \eqref{eq:10dsusy1}, we find further differential conditions, under the assumption that we solve $m q_-=0$  as $q_-=0$ (necessary for AdS$_5$). In summary, supersymmetric AdS$_5$ vacua must satisfy
\begin{subequations} 
\begin{align}
&e^{-2A}q_+=c,~~~~\nabla_{(a}\xi_{b)}=0,~~~~d\tilde\xi=\iota_{\xi}H\label{eq:intsusy00ads5},\\[2mm]
&d_H(e^{2A-\Phi}\psi^2_+)=0,~~~~d_H(e^{3A-\Phi}\psi^2_-)-3m ie^{2A-\Phi}\psi^2_+=0,\label{eq:intsusy1ads5}\\[2mm]
&d_H(e^{3A-\Phi} \text{Re}\psi^1_-)=0,~~~~d_H(e^{A-\Phi} \text{Im}\psi^1_-)=0,\label{eq:intsusy2ads5}\\[2mm]
&d_H(e^{2A-\Phi}\text{Re}\psi^1_+)+2m e^{A-\Phi}\text{Im}\psi^1_-=0,\label{eq:intsusy3ads5}\\[2mm]
&d_H(e^{4A-\Phi}\text{Im}\psi^1_+)-4m e^{3A-\Phi}\text{Re}\psi^1_--\frac{c}{4}e^{5A}\star \lambda(f_+)=0,\label{eq:intsusy4ads5}\\[2mm]
&d_H(e^{-\Phi} \text{Im}\psi^1_+)+\frac{1}{4}(\iota_{\xi}+\tilde{\xi}\wedge) f_+=0,~~~~d_H(e^{5A-\Phi}\text{Im}\psi^1_-)+\frac{1}{4}e^{5A}(\iota_{\xi}+\tilde{\xi}\wedge)\star \lambda(f_+)=0.\label{eq:intsusy7}
\end{align}
\end{subequations}
These conditions are necessary and sufficient for AdS$_5$ vacua. The conditions \eqref{eq:intsusy7} when extracted are multiplied by $f$, however since they are implied by the rest of the conditions we present, irrespective of the value of $f$ we can remove the $f$ multiplicative factor. They will be important for the embedding of the $d=5$ minimal supergravity. Note that when one fixes $m=0$ we also have conditions for Mink$_5$ vacua, though not completely general ones which do not demand $q_-=0$ --- however this constraint is necessary if one wishes to allow for purely RR sources.

\subsection{\texorpdfstring{AdS$_5$}{AdS(5)} vacua}\label{sec:onAdS5vaccua}
Following \cite{Apruzzi:2015zna} one solves the bi-spinor constraints of the previous section by first decomposing the internal spinors in a common basis in terms of a single spinor $\chi$ with norm $||\chi||^2=e^{A}c$. This leads to the bi-spinors
\beq
\chi\otimes \chi^{\dag}= \frac{e^{A}c}{4}(1+v)\wedge e^{-i j_2},~~~~\chi\otimes \chi^{c\dag}=\frac{e^{A}c}{4}(1+v)\wedge \omega_2,
\eeq
where $(v,j_2,\omega_2)$ span an SU(2)-structure on M$_5$. Consistency with \eqref{eq:zeroforms} and the 0-form part of the second of \eqref{eq:intsusy1ads5} restricts this decomposition to
\beq\label{eq:spinordecomposition}
\chi_1= \chi,~~~~\chi_2= a \chi+ \frac{b}{2}\overline{w}\chi,~~~~a=a_1+i a_2~~~~a_1^2+a_2^2+b^2=1,
\eeq
for $w$ a holomorphic 1-form such that $||w||^2=2$, $\iota_v w=0$ and $w\chi=0$. Defining a second 1-form as $z=-\frac{1}{2}\iota_w \omega_2$, the SU(2)-structure forms then decompose as\footnote{With respect to \cite{Apruzzi:2015zna} we have a sign change in $\omega_2$. This is due to a difference in phase in the internal intertwiner defining Majorana conjugation (see appendix \ref{sec:appendix}). The choice we make here ensures that \eqref{eq:internalbilinears2}-\eqref{eq:internalbilinears3} takes the same form as \cite{Apruzzi:2015zna} (up to the $e^A c$ factor explained in the next footnote), which is what actually matters if we wish to use their results.}
\beq
j_2= \frac{i}{2}(w\wedge \overline{w}+z\wedge \overline{z}),~~~~\omega_2= w\wedge z,
\eeq
with $\{v,\text{Re}w,\text{Im}w,\text{Re}z,\text{Im}z\}$ giving a vielbein on M$_5$. The internal bi-linears of the previous section then decompose in terms of this vielbein as
\begin{subequations}
\begin{align}
\xi&= e^{A}c b(b v-\text{Re}(a w)),~~~~&\tilde{\xi}&=e^{A}c (b \text{Re}(a w)+(1-b^2) v),\label{eq:internalbilinears1}\\[2mm]
\psi^1_+&=\frac{e^{A} c}{4}\overline{a}e^{-i j_2+\frac{b}{\overline{a}}v\wedge w},~~~&\psi^1_-&= \frac{e^{A} c}{4}(\overline{a} v+b w) \wedge e^{-i j_2},\label{eq:internalbilinears2}\\[2mm]
\psi^2_+&=\frac{e^{A} c}{4}(a w-b v)\wedge z\wedge e^{-i j_2},~~~~~&\psi^2_-&=-\frac{e^{A}c}{4}b z\wedge e^{-i j_2- \frac{a}{b} v\wedge w}\label{eq:internalbilinears3}
\end{align}
\end{subequations}
which span an identity-structure.\footnote{In \cite{Apruzzi:2015zna} $c=1$, which one can choose to fix without loss of generality. The $e^{A}$ factor has been extracted appearing instead in \eqref{eq:intsusy00ads5}-\eqref{eq:intsusy4ads5}.} The condition that $\xi$ is Killing allows us to parameterise it as
\beq
\xi= \frac{||\xi||^2}{3c}D\psi,~~~~~ D\psi\equiv d\psi+ V,~~~~||\xi||= b e^{A}c,
\eeq
with $\partial_{\psi}$ a Killing vector and $V$ a 1-form with support on the directions of M$_5$ that are not $\psi$. We then have that M$_5$ decomposes as a U(1) fibration over a four-dimensional base as
\beq
ds^2(\text{M}_5)= \frac{||\xi||^2}{9c^2}D\psi^2+ ds^2(\text{M}_4),
\eeq
with M$_4$ independent of $\psi$ (at least locally). \\
~\\
Introducing coordinates $(s, u, x_1, x_2)$ on M$_4$, the problem of finding supersymmetric AdS$_5$ solutions can be recast in terms of two functions $(D_u, D_s)$ depending on $(s, u, x_1, x_2)$, subject to partial differential equations \cite{Bah:2017wxp}. 

The metric for a general supersymmetric AdS$_5$ solution is 
\begin{subequations}
\begin{align}
ds^2_{10} &= e^{2A} \left[ ds^2({\rm AdS}_5) + e^{2\varphi} \left(dx_1^2 + dx_2^2 \right) + \frac{1}{3} e^{-6\lambda} ds^2_3 \right]\, , \label{5-5metric} \\ 
ds^2_{3} &= -\frac{4}{\partial_s D_s} D\psi^2 - \partial_s \widetilde{D}_s\, ds^2  - 2 \partial_u D_s\, du ds  - \partial_u D_u\, du^2\, , \label{eq:3metric}
\end{align}
\end{subequations} 
where
\begin{equation}
D\psi = d\psi -\frac{1}{2 m} \star_2 d_2 D_s
\end{equation}
and $\widetilde{D}_s \equiv D_s - \frac{3}{2}\ln s$.
The Hodge star operator $\star_2$\footnote{The convention for its action is $\star_2 dx_1 =dx_2$ and $\star_2 dx_2 =-dx_1$.} and the exterior derivative $d_2$ are taken over the $(x_1,x_2)$ plane. 

The functions appearing in the metric are given in terms of $(D_u, D_s)$ as follows:
\begin{equation}
e^{-6\lambda} = \frac{1}{8 m^2s} \frac{\det(h)}{\det(g)}\,, \qquad e^{4A} = - \frac{\partial_s D_s}{3\det(h)}\,, \qquad e^{2\varphi} = \frac{1}{24 m^2} \det(h) e^{D_s}\,,
\end{equation} 
with
\begin{equation}
\begin{split}
	\det(g) &=\partial_u D_u \partial_s \widetilde{D}_s - \left(\partial_u D_s \right)^2\,, \\
	\det(h) &=\partial_u D_u \partial_s D_s - \left(\partial_u D_s \right)^2\,.  
\end{split}	
\end{equation}
The dilaton can be expressed as
\begin{equation}
e^{2\Phi} =  e^{6A} e^{-6\lambda}\,.
\end{equation}
The NSNS field strength $H$ is given by 
\begin{align}
H&= \frac{1}{3c}d\bigg[\tilde\xi\wedge D\psi+\frac{c}{8m^2\sqrt{2s}}\partial_u(e^{D_s})dx_1\wedge dx_2\bigg]+\frac{1}{36m^2\sqrt{2s}}du\wedge d \star_2 d_2 D_s- \frac{e^{D_s}}{12 cm} \det(g)\tilde\xi\wedge dx_1\wedge dx_2,\nn\\[2mm]
\tilde\xi&=-\frac{c}{6m \det(g)\sqrt{2s}}\left(\frac{3}{2s}\partial_uD_sds + \det(h)du\right)\label{eq:Hform}
\end{align}  
The RR field strengths read
\begin{align}
F_0 &= 36 \sqrt{2s}m^2 \frac{\partial_u \left(\partial_s D_u - \partial_u D_s \right)}{\partial_s D_s}\,  \label{eq:F0},\\[2mm]
F_2&= \frac{1}{3c}F_0\tilde\xi\wedge D\psi-d\left(\star_2 d_2 D_u+2m \frac{\partial_uD_s}{\partial_sD_s}D\psi\right)\nn\\[2mm]
&+\left(\Delta_2D_u-\partial_u(e^{D_s}s \det(g))\right)dx_1\wedge dx_2+\star_2 d_2\left(\partial_uD_s-\partial_sD_u\right)\wedge ds,\label{eq:F2}\\[2mm]
F_4&= \frac{1}{3c}F_2\wedge\tilde\xi\wedge D\psi-\frac{1}{36 m}d\left(\sqrt{2 s}e^{D_s}\det(h)dx_1\wedge dx_2\wedge D\psi\right)+\frac{1}{18m \sqrt{2s}}ds\wedge d(\star_2 d_2D_s)\wedge D\psi\nn\\[2mm]
&+\frac{\partial_uD_s}{18m\sqrt{2s}\partial_sD_s}\bigg[du\wedge \bigg(d(\star_2 d_2 D_s)+\frac{1}{2}e^{D_s}\det(h)dx_1\wedge dx_2\bigg)+\frac{3}{2}d\left(\partial_u(e^{D_s})\right)\wedge dx_1\wedge dx_2\bigg]\wedge D\psi,\label{eq:F4}
\end{align}
where $\Delta_2$ is the Laplace operator $\Delta_2 = \partial^2_{x_1} + \partial^2_{x_2}$.

The Bianchi identity of the Romans mass $F_0$ sets it to a constant.  The Bianchi identity of the NSNS field strength, $dH = 0$,  yields an equation for $D_s$:
\begin{equation}\label{eq:dH}
\Delta_2 D_s = \partial_s \left(s \det(g) e^{D_s} \right) + \frac{1}{24 m^2\sqrt{2s}} F_0 \partial_s e^{D_s}\, ,
\end{equation} 
which is actually also required for supersymmetry to hold, so there can be no NSNS sources.. 
Given the above, it follows that the Bianchi identity of $F_2$, $dF_2 - F_0 H = 0$, is equivalent to
\begin{equation}\label{eq:dF2}
\Delta_2\left(\partial_u D_u\right) = \partial_u^2 \left(s \det(g) e^{D_s} \right)  + \frac{1}{36 m^2\sqrt{2s}} F_0 s \partial_s \left(\det(h) e^{D_s} \right)\, . 
\end{equation}
The Bianchi identity of $F_4$ is automatically satisfied.\\
~~\\
A general class of solutions contained within this framework are the AdS$_7$ holographic duals of six-dimensional, $\mathcal{N}=(1,0)$ theories studied in \cite{Apruzzi:2013yva}, compactified on a Riemann surface, giving rise to four-dimensional, $\mathcal{N}=1$ SCFTs and their anti-de Sitter duals. Examples of AdS$_5$ solutions arising from compactifications on a Riemann surface with genus $g>1$ were studied in \cite{Apruzzi:2015zna, Apruzzi:2015wna}. The extension to punctured Riemann surfaces was studied in \cite{Bah:2017wxp}. Another class of solutions which may be embedded in the above classification, with vanishing Romans mass, are the abelian and non-abelian T-duals of the Sasaki--Einstein solutions, though the details of the explicit embedding have not been worked out fully. Indeed consistent truncations on the non-abelian T-duals of $S^5$, $T^{1,1}$ and $Y^{p,q}$ to $d=5$ minimal gauged supergravity were constructed in \cite{Cheung:2022wpg} recently and can be seen as a particular choice of background of the general formalism we present in the following sections.

\subsection{\texorpdfstring{Mink$_5$}{Mink(5)} vacua}\label{sec:onmink5vaccua}
In this section we shall present a sub-class of possible Mink$_5$ vacua, namely those consistent with the spinor ansatz \eqref{eq:spinordecomposition} with $b\neq 0$ that can be used to embed $d=5$ minimal ungauged supergravity into massive Type IIA supergravity. To our knowledge these do not appear anywhere else in the literature.\\
~~\\
It is possible to show that when $m=0$ \eqref{eq:intsusy00ads5}-\eqref{eq:intsusy3ads5} can be solved  in terms of local coordinates $(\psi,x_1,x_2,s,u)$ and the vielbein
\begin{align}
v&= e^{A}\left(\frac{b^2}{3}D\psi+ e^{-5A+\Phi}(a_1 du-a_2 e^{2A+k} ds)\right),~~~~z=- i e^{-4A+\Phi}b^{-1}(dx_1+ i dx_2),\nn\\[2mm]
w&=b^{-1}\left(-\overline{a}\left(\frac{e^Ab^2}{3}D\psi-i e^{-4A+\Phi}(a_2 du+e^{2A+k}a_1ds)\right)+ e^{-4A+\Phi}b^2(du+i e^{2A+k}ds)\right),
\end{align}
where $e^{k}$ is a function of $s$ only, and parametrises diffeomorphism invariance in this direction and
\beq
a=e^{-\Phi}(c_0p^{-\frac{3}{4}} (q-l^2)^{\frac{3}{4}}+i p^{-\frac{5}{4}} l (q-l^2)^{\frac{1}{4}}),~~~~b=e^{-\Phi}p^{-\frac{5}{4}}(q-l^2)^{\frac{3}{4}},~~~e^A= p^{\frac{1}{4}}(q-l^2)^{-\frac{1}{4}},
\eeq
where $c_0$ is a constant and the constraint $|a|^2+b^2=1$ fixes $e^{-\Phi}$. Here $p$ has support on $(s,x_1,x_2)$ and $(q,l)$ on $(u,x_1,x_2)$. The connection appearing in $D\psi$ is fixed such that
\beq
dV= -3du\wedge \star_2 d_2 l+ 3\partial_u l dx_1\wedge dx_2-3 c_0 e^{k}ds\wedge \star_2 d_2 p,
\eeq
where for consistency with $d^2V=0$ we should have
\beq
c_0 \Delta_2 p=0,~~~(\partial_u^2+\Delta_2 )l=0,
\eeq
where $\Delta_{2}$ is again flat space Laplacian on $(x_1,x_2)$.
What remains non-trivial in  \eqref{eq:intsusy00ads5}-\eqref{eq:intsusy3ads5} fixes the NSNS flux; we find 
\beq
H=\frac{1}{3c}\left(D\psi\wedge d\tilde\xi+d( \tilde{\xi}\wedge V)\right)+ e^{k} ds\wedge du\wedge \star_2 d_2 p.
\eeq
What remains to solve is \eqref{eq:intsusy4ads5}-\eqref{eq:intsusy7}, which simply define the RR fluxes. We shall quote them along with our summary of the class.\\
~~\\
In summary, we find a class of Mink$_5$ vacua with NSNS sector of the form
\begin{align}
ds^2_{10}&=\sqrt{p}\bigg[\frac{1}{\sqrt{\Xi_1}}ds^2(\text{Mink}_5)+ \sqrt{\Xi_1}\bigg(\frac{1}{\Xi_2}\bigg(\frac{1}{9 q}D\psi^2+ Du^2\bigg)+ e^{2k}\frac{p}{q}ds^2+ dx_1^2+dx_2^2\bigg)\bigg],\nn\\[2mm]
H&= \frac{1}{3c}d\left(\tilde{\xi}\wedge D\psi\right)+ e^{k} ds\wedge du\wedge\star_2 d_2 p,~~~~e^{-\Phi}=p^{-\frac{5}{4}}\Xi_1^{\frac{1}{4}}\sqrt{q \Xi_2}\label{eqminkNS}
\end{align}
where we define
\beq
\Xi_1\equiv (q-l^2),~~~ \Xi_2\equiv 1+ \frac{p}{q}c_0^2\Xi_1,~~~\tilde{\xi}=\frac{c p}{q\Xi_2}(c_0 \Xi_1 du-l e^{k}ds),~~~Du\equiv du+c_0\frac{pl}{q} e^{k}ds.
\eeq
These backgrounds support several RR fluxes, which can be compactly expressed in terms of
\beq
B= \frac{1}{3c}\tilde\xi\wedge D\psi-e^k u ds\wedge\star_2 d_2 p,
\eeq
which away from NSNS sources is such that $dB=H$; we find
\begin{align}
F_0&=\frac{1}{p}\left(c_0 \partial_u l-e^{-k} q \partial_s(p^{-1})\right),\nn\\[2mm]
F_2&=F_0 B-\frac{1}{3}d\left(\frac{l}{p}D\psi\right)-du\wedge \star_2 d_2(qp^{-1})+p^{-1}\partial_u\left(q+c_0^2p\Xi_1\right)dx_1\wedge dx_2+e^k ds\wedge \left(F_0u \star_2 d_2p-c_0 \star_2 d_2l\right),\nn\\[2mm]
F_4&=B\wedge F_2-\frac{1}{2}B\wedge B F_0- \frac{c_0}{3}d\left(\Xi_1D\psi\right)\wedge dx_1\wedge dx_2+\frac{1}{9}e^{k}ds\wedge dV\wedge D\psi\nn\\[2mm]
&-\frac{1}{3}e^{k}ds\wedge\bigg[ d\left(u l p^{-1}D\psi\right)+3 u du\wedge \star_2 d_2(qp^{-1})\bigg]\wedge \star_2 d_2 p \label{eqminkRR}. 
\end{align}
Away from the loci of sources the Bianchi identities of the RR and NSNS fluxes demand $F_0$ is constant and that the following partial differential equations are solved,
\begin{align}
\Delta_2 p=0,~~~~(\partial_u^2+\Delta_2 )l=0,~~~~\Delta_2(q p^{-1})+ \partial_u^2\left(q+c_0^2 p \Xi_1\right)=0
\end{align}
which define solutions in this class. The second of these is implied by $d^2V=0$, while the first is also implied by this when $c_0\neq 0$. We remind the reader that $\partial_u p= \partial_s q=\partial_s l=0$, so arranging for $F_0=$ constant and solving the final partial differential equation leads to branching  classes of solutions, the most obvious are those for which either of the $s$ or $u$ directions become an isometry direction.\\
~~\\
To our knowledge this is the first time this class of Mink$_5$ solutions has appeared in the literature, a detailed analysis of the solutions it contains is outside the scope of this work. We note however that constructing compact solutions is not particularly difficult: The simplest non-trivial solution is probably given by fixing
\beq
c_0=l=0,~~~~ q=1,~~~~ p=e^{-k}=h_8^{-1},
\eeq
for $h_8=h_8(x)$. This reduces the class to the solution of formal D8-branes along Mink$_5\times \mathbb{T}^4$  with $h_8$ locally a linear function and $\partial_x h_8=F_0$. Locally this makes $x$ span a semi-infinite interval bounded at one end by an O8--D8 system --- globally however one can glue such local patches together with D8-branes in the fashion of \cite{Macpherson:2018mif} (see section 4.1 therein) thereby bounding $x$ between two O8--D8 singularities with additional D8-branes along the interior.

\section{Embedding of $d=5$ minimal (un)gauged supergravity into massive Type IIA supergravity}\label{eq:theembedding}
In this section we will embed $d=5$ minimal (un)gauged supergravity into massive Type IIA supergravity. The action of the bosonic part of this theory, in mostly positive metric conventions, is
\begin{equation}
S=\int \bigg[ \big(R^{(5)}+12 m^2 \big)\star_5 1 -\frac{1}{6} {\cal F}\wedge \star_5{\cal F}-\frac{1}{27} {\cal A}\wedge {\cal F}\wedge {\cal F}\bigg],
\end{equation}
where ${\cal F}=d{\cal A}$. The equations of motion following from the action are
\begin{subequations}
\begin{align}
&R^{(5)}_{\mu\nu}=-4 m^2 g^{(5)}_{\mu\nu}+\frac{1}{6}{\cal F}_{\mu\rho}{\cal F}_{\nu}^{~\rho}-\frac{1}{36} g^{(5)}_{\mu\nu}{\cal F}_{\rho\sigma}{\cal F}^{\rho\sigma}\, ,\label{eq:exein}\\
&d\star_5 {\cal F}+\frac{1}{3} {\cal F}\wedge {\cal F}=0\, ,\label{eq:exfluxeom}
\end{align}
\end{subequations}
while the preservation of supersymmetry requires the vanishing of the gravitino variation which implies
\begin{equation}
 \bigg[\nabla_\mu +\frac{ m}{2} \mathcal{A}_{\mu}-\frac{m}{2} \gamma_{\mu}+\frac{1}{24}\mathcal{F}_{\rho\sigma}\Big(\gamma{_{\mu}^{~\rho\sigma}}-4 \delta_{\mu}^{\rho}\gamma^{\sigma}\Big)\bigg]\zeta\,=0.\label{eq:externalSUSY}
\end{equation}
Notice that when ${\cal A}=0$ this reduces to the Killing spinor equation of AdS$_5$, and the equations of motion reduce to $R_{\mu\nu}=-4m^2 g_{\mu\nu}$ making AdS$_5$ the vacuum of this theory, at least for $m\neq0$. Solutions of minimal gauged supergravity were classified in \cite{Gauntlett:2003fk}, and in the ungauged limit $m=0$ in \cite{Gauntlett:2002nw}.\\
~~\\
One can embed $d=5$ minimal supergravity into $d=10$ by again taking the spinor ansatz \eqref{eq:spinoransatz}, with $\zeta$ now taken to obey \eqref{eq:externalSUSY}. The $d=10$  bi-linears decompose in the same fashion as they do in \eqref{eq:10dbilinears} for AdS$_5$ vacua, albeit now for generalised $d=5$ bi-spinors $\phi^{1,2}$. As the external spinor now obeys \eqref{eq:externalSUSY}, clearly \eqref{eq:5ddiffbi} are no longer valid, indeed one can show these are modified to
\begin{subequations}
\begin{align}
&\nabla_{(\mu}k_{\nu)}=0,\\[2mm]
d\phi^1_-&=m (\text{deg})\phi^1_+ +\frac{2i}{3(\text{deg}!)}\phi_+^1\wedge {\cal F}-\frac{1}{12}\iota_k\star_5 {\cal F},~~~~d\phi^1_+= -\frac{i}{12}\iota_k {\cal F},\label{eq:gen5dconds1}\\[2mm]
(d+i m{\cal A}\wedge )\phi^2_2&=3m \phi^2_3,~~~~(d+i m{\cal A}\wedge )\phi^2_3=\frac{i}{3}{\cal F}\wedge \phi^2_2\label{eq:gen5dconds2}.
\end{align}
\end{subequations}
We again define $(f,k)$ as in \eqref{eq:somedefs} (note that $f$ is no longer necessarily constant), it then follows from the differential bi-spinor relations that
\beq\label{eq:liebispinors}
{\cal L}_k\phi^1=0,~~~{\cal L}_k\phi^2= im(3 f-\iota_k{\cal A})\phi^2,~~~~d{\cal F}=0~~\Rightarrow~~ {\cal L}_k {\cal F}=0,
\eeq
which makes $k^{\mu}\partial_{\mu}$ a Killing vector under which $\zeta$ is charged, as it was for AdS$_5$. It then follows again that  $\xi^a\partial_a$ must be a Killing vector for  $\nabla_{(M}K^{(10)}_{N)}=0$ to hold.\\
~~ \\
We seek an embedding of $d=5$ minimal supergravity that, like the Type IIB and M-theory examples \cite{Gauntlett:2007ma}, does not ultimately require supersymmety to hold. As such the $d=10$ bosonic fields of massive Type IIA supergravity should only depend on those of the $d=5$ theory, and not $\phi^{1,2}$ which require a Killing spinor to define. We shall assume that, like for AdS$_5$, $\xi$ does not vanish so we have a U(1) isometry which can be fibred over the $d=5$ supergravity directions by ${\cal A}$. We thus take the ten-dimensional metric to be\footnote{The precise numerical factor multiplying ${\cal A}$ can be fixed in several ways, perhaps the quickest is consistency with the fact that $(K^{(10)})^M\partial_M$ should be a Killing vector under the assumption that $\partial_{\psi}$ is itself Killing and given that $k^{\mu}\partial_{\mu}$ is Killing.}
\beq
ds^2_{10}=e^{2A} g^{(5)}_{\mu\nu}dx^{\mu}dx^{\nu}+  \frac{\xi^2}{||\xi||^2}+ ds^2(\text{M}_4),~~~~ \frac{\xi}{||\xi||}= \frac{||\xi||}{3c}{\cal D}\psi,~~~~{\cal D}\psi\equiv d\psi+V-{\cal A},
\eeq
where $e^{2A}$ and the dilaton have support on M$_4$ alone. Consequently we have
\beq
16(K^{10})^M\partial_M=  e^{-2A}q_+ (k^{\mu}\partial_{\mu}+\iota_k {\cal A}\partial_{\psi})-3e^{-2A}q_+ f \partial_{\psi},~~~~\xi^a\partial_a =  3 c \partial_{\psi}.
\eeq
In addition to imposing that $\xi^a\partial_a$ is a Killing vector \eqref{eq:10dsusy2} demands, for the NSNS 3-form to be independent of $\phi^{1,2}$, that
\beq
e^{-2A}q_+=c,~~~~q_-=0,~~~~ H=H_3+ \frac{1}{3c}({\cal D}\psi\wedge d\tilde\xi+\tilde{\xi}\wedge {\cal F}),\label{eq:firstfixing}
\eeq
for $c>0$ a constant and where $H_3$ is orthogonal to both ${\cal D}\psi$ and the external directions --- notice that $d{\cal F}=0$ implies that $dH$ is independent of external data. The first of  \eqref{eq:decendent}, given \eqref{eq:liebispinors}, then furnishes us with information about the charge of the internal bi-spinors under $\partial_{\psi}$, namely
\beq
{\cal L}_{\xi}\psi^1=0,~~~~ {\cal L}_{\xi}\psi^2= 3 im c \psi^2,
\eeq
meaning that one can locally take the only functional dependence of $\psi$ in these bi-spinors to be  an $e^{im \psi}$ factor in $\psi^2$.
To proceed we demand that the RR fluxes can close on the Bianchi identity and equation of motion of ${\cal F}$ which means it can only depend on external data through $({\cal F},\star_5 {\cal F},{\cal D}\psi)$ restricting its form to\footnote{The ${\cal D}\psi$ terms are contained implicitly in $(g_+,f_+)$. One might think of including all the combinations one can construct out of $({\cal A},{\cal F})$, utilising hodge duals and wedge products, however there are no necessary external conditions which these close on. For instance ${\cal F}\wedge {\cal F}$ at first sight may appear reasonable to include, but the self-duality constraint $F$ must obey means this must come with $\star_5({\cal F}\wedge {\cal F}$), which need obey no special identity under $d$.}
\beq
F= f_++e^{2A}{\cal F}\wedge g_+-e^{3A}\star_{5} {\cal F}\wedge \star \lambda(g_+) +e^{5A}\text{vol}_5\wedge \star\lambda(f_+),
\eeq
where $(f_+,g_+)$ are defined on $({\cal D}\psi,\text{M}_4)$ and are to be determined. We are now ready to reduce  \eqref{eq:10dsusy1} to conditions on the internal space. To deal with the fact that ${\cal D}\psi$ also contains the external potential ${\cal A}$, one can decompose all the objects defined on M$_5$ into their parts defined along the base and fibre directions, i.e.
\beq
\psi^1= \psi^{1B}+ {\cal D}\psi\wedge \psi^{1F},~~~~\psi^2=e^{i m\psi}(\psi^{2B}+ {\cal D}\psi\wedge \psi^{2F}),\label{eq:base fiber}
\eeq
and so on. One then again substitutes for $\Psi^{(10)}$ in \eqref{eq:10dsusy1}, this time making use of \eqref{eq:gen5dconds1}-\eqref{eq:gen5dconds2} and attempts to factor out the external data. Since we want to embed all solutions of $d=5$ minimal supergravity in a common framework, there are no identities we can assume that the wedge products of the external fields and bi-spinors obey, i.e.\  we must take $({\cal F}\wedge \phi^1)_n$ to be independent of $\phi^1_{n+2}$ and so on. However there are important identities that the internal bi-spinors obey, namely\footnote{These follow from the necessary $d=10$ condition $(\iota_{K^{(10)}}+\tilde{K}^{(10)}\wedge)\Psi^{(10)}=0$ given \eqref{eq:firstfixing} and  the identities involving $\phi^{1,2}$ in appendix \ref{eq:deq5lordetails}
.}
\beq\label{eq:internalidentities}
(\iota_{\xi}+\tilde{\xi}\wedge )\psi^{1,2}_{+}= e^{A}c \psi^{1,2}_{-},~~~~(\iota_{\xi}+\tilde{\xi}\wedge )\psi^{1,2}_{-}=0,
\eeq
which can in turn be decomposed into conditions on base and fiber directions as in \eqref{eq:base fiber}. Putting this all together, after a lengthy computation,  we find that \eqref{eq:10dsusy1} reduces to a number of conditions on the internal space we can most succinctly express as
\begin{subequations}
\begin{align}
&e^{2A}g_+=-\frac{4}{3 c} e^{-\Phi} \text{Im}\psi^1_+\label{eq:intsusy111},\\[2mm]
&d_H(e^{2A-\Phi}\psi^2_+)\bigg\lvert_{{\cal A}=0}=0,~~~~d_H(e^{3A-\Phi}\psi^2_-)-3m ie^{2A-\Phi}\psi^2_+\bigg\lvert_{{\cal A}=0}=0,\label{eq:intsusy1ads5s}\\[2mm]
&d_H(e^{3A-\Phi} \text{Re}\psi^1_-)\bigg\lvert_{{\cal A}=0}=0,~~~~d_H(e^{A-\Phi} \text{Im}\psi^1_-)\bigg\lvert_{{\cal A}=0}=0,\label{eq:intsusy2ads5s}\\[2mm]
&d_H(e^{2A-\Phi}\text{Re}\psi^1_+)+2m e^{A-\Phi}\text{Im}\psi^1_-\bigg\lvert_{{\cal A}=0}=0,\label{eq:intsusy3ads5s}\\[2mm]
&d_H(e^{4A-\Phi}\text{Im}\psi^1_+)-4m e^{3A-\Phi}\text{Re}\psi^1_--\frac{c}{4}e^{5A}\star \lambda(f_+)\bigg\lvert_{{\cal A}=0}=0,\label{eq:intsusy4ads5s}\\[2mm]
&d_H(e^{-\Phi} \text{Im}\psi^1_+)+\frac{1}{4}(\iota_{\xi}+\tilde{\xi}\wedge) f_+\bigg\lvert_{{\cal A}=0}=0,~~~~d_H(e^{5A-\Phi}\text{Im}\psi^1_-)+\frac{1}{4}e^{5A}(\iota_{\xi}+\tilde{\xi}\wedge)\star \lambda(f_+)\bigg\lvert_{{\cal A}=0}=0,\label{eq:intsusy7s}
\end{align}
\end{subequations}
where we prove these conditions are indeed necessary and sufficient for supersymmetry in appendix \ref{eq:susysuff}. We have now reproduced \eqref{eq:intsusy00ads5}-\eqref{eq:intsusy7} and therefore the internal space of the embedding of $d=5$ minimal gauged supergravity is mostly the same as it is for the AdS$_5$ vacua, with the only modifications happening in the fluxes and U(1) fiber in terms of $({\cal A},{\cal F})$. In particular, away from the loci of sources, we necessarily have  that   
\beq
dH\bigg\lvert_{{\cal A}=0}=0,~~~~ d_H F\bigg\lvert_{{\cal A}=0}=0,
\eeq
as the left-hand side of these expressions reduce to their AdS$_5$ vacua values.\footnote{Strictly speaking the form of vol$_5$ depends on the specific solution of $d=5$ minimal supergravity, but the parts of $d_H F$ parallel and orthogonal to this vanish independently, so this subtlety is immaterial.} It is then a simple matter to show that the NSNS Bianchi identity is implied by $d{\cal F}=0$ and with a little more effort that
\beq
d_H F=d_H F\bigg\lvert_{{\cal A}=0}
\eeq
is implied by \eqref{eq:intsusy111}-\eqref{eq:intsusy7s}, the identities \eqref{eq:internalidentities} and the Bianchi identity and equation of motion of ${\cal F}$. Therefore the Bianchi identities of the fluxes are implied by the AdS$_5$ result. We prove in appendix \ref{eq:EOMsuff} that all the equations of motion of Type IIA supergravity are implied by what we present in this section, irrespective of whether the solution on the external space is supersymmetric or not.\\
~~\\
In summary the embedding of $d=5$ minimal supergravity into massive Type IIA supergravity is given by
\begin{align}
ds^2_{10}&=e^{2A} g^{(5)}_{\mu\nu}dx^{\mu}dx^{\nu}+ \frac{||\xi||^2}{9c^2}{\cal D}\psi^2+ ds^2(\text{M}_4),~~~~{\cal D}\psi\equiv d\psi+V-{\cal A}\nn\\[2mm]
H&=H_3+ \frac{1}{3c}({\cal D}\psi\wedge d\tilde\xi+\tilde{\xi}\wedge {\cal F}),\nn\\[2mm]
F&= f_++e^{5A}\text{vol}_5\wedge \star\lambda(f_+)-\frac{4}{3 c}\bigg[{\cal F}\wedge  (e^{-\Phi} \text{Im}\psi^1_+)-\star_{5} {\cal F}\wedge (e^{A-\Phi} \text{Im}\psi^1_-)\bigg]\label{eq:embeddign}.
\end{align}
More specifically $d=5$ minimal gauged supergravity can be embedded into massive Type IIA supergravity by making the following substitutions\footnote{Note that $D\psi\to {\cal D}\psi= D\psi-{\cal A}$ must be substituted before evaluating $dD\psi$ where it appears.} in the AdS$_5$ vacua of \eqref{5-5metric}-\eqref{eq:F4}
\begin{align}
&ds^2(\text{AdS}_5)\to g^{(5)}_{\mu\nu}dx^{\mu}dx^{\nu},~~~~~\text{vol}(\text{AdS}_5)\to  \text{vol}_5,\nn\\[2mm]
&D\psi\to {\cal D}\psi= D\psi-{\cal A},~~~~F_4 \to F_4-\frac{1}{3 \sqrt{2s}} \left(-\frac{1}{3}{\cal F}\wedge {\cal D}\psi+\star_5 {\cal F}\right)\wedge ds.\label{eq:embeddinggauged}
\end{align}
The ungauged limit, $m=0$, on the other hand can be embedded into massive Type IIA supergravity by making the following substitutions in the Mink$_5$ vacua of \eqref{eqminkNS}- \eqref{eqminkRR}
\begin{align}
&ds^2(\text{Mink}_5)\to g^{(5)}_{\mu\nu}dx^{\mu}dx^{\nu},~~~~~\text{vol}(\text{Mink}_5)\to  \text{vol}_5,\nn\\[2mm]
&D\psi\to {\cal D}\psi= D\psi-{\cal A},~~~~F_4 \to F_4+ \frac{e^{k}}{3} \left(-\frac{1}{3}{\cal F}\wedge {\cal D}\psi+\star_5 {\cal F}\right)\wedge ds.
\end{align}

\section{Uplift examples}\label{sec:examples}

In this section we will use the consistent truncation constructed in the previous section to uplift two families of solutions of $d = 5$ minimal gauged supergravity to massive Type IIA supergravity. As seed AdS$_5$ solution on which to perform the uplift we take the infinite family of solutions constructed in \cite{Bah:2017wxp} (BPT). Another interesting class of solutions which we could use as our seed solutions are the ones found in \cite{Apruzzi:2015zna}. The field theory duals of these solutions, along with the interpolating flow between the AdS$_7$ and AdS$_5$ vacua have recently been studied in \cite{Merrikin:2022yho}; see also \cite{Faedo:2019cvr}. One could then use these solutions to construct supergravity solutions dual to the compactification of a six-dimensional, $\mathcal{N}=(1,0)$ theory down to, for example, a two-dimensional SCFT, in a two stage process. This should, in theory, allow for greater control in understanding the two-dimensional SCFT through this chain of reductions rather than performing a direct reduction on a four-dimensional space. It would thus be interesting to further compactify the quiver theories studied there, the holographic duals of which will be accessible by using our truncation.
We begin by reviewing the seed AdS$_5$ solutions, before rewriting them in a form consistent with the uplift formula presented in section \ref{eq:theembedding}. We then review the two classes of solutions of $d = 5$ minimal gauged supergravity that we will uplift, before studying some basic properties of the uplifted solutions.

\subsection{Reduction on solutions of BPT}\label{sec:uplift}

In this section we will review the solutions found in \cite{Bah:2017wxp} before constructing solutions using the truncation discussed above.\footnote{We will set $c=1$ and $m=1$ in this section to avoid cluttering the equations.}
The solutions of \cite{Bah:2017wxp} have the broad interpretation of placing a six-dimensional, $\mathcal{N}=(1,0)$ theory on a constant curvature Riemann surface of genus $g$, though different completions allow for different interesting physics. The metric takes the form
\begin{align}
\dd s^2_{10}&=\me^{2 A}\bigg[\dd s^2({\text{AdS}_5})-\frac{p'(z)}{9 z^2}\dd s^2(X_5)\bigg]\, ,\\
\dd s^2(X_5)&= \dd s^2 (\Sigma_g)+\frac{3 z \dd z^2}{p(z)}+\frac{9 z^3}{3 p(z)-z p'(z)}\bigg[\frac{k}{1-k^3} \dd k^2+\frac{4(1-k^3)p(z)}{3\big(3 p(z)-z p'(z)(1-k^3)\big)}D\psi^2\bigg]\, ,
\end{align}
where $p(z)$ is the cubic function
\be
p(z)=(z-z_0)\big(\kappa(z^2+z_0 z+z_0^2) -3 \ell z_1^2\big)\,,\qquad \Rightarrow \qquad p'(z)=3(\kappa z^2- \ell z_1^2)\, , 
\ee
and 
\be
D\psi\equiv \dd \psi - A_g\, ,\quad \dd A_g=\vol(\Sigma_g)\, .
\ee
The parameters $\kappa$, $z_0$, $z_1$ and $\ell$ are all real, with $\kappa = 0, \pm 1$ and $\ell$ constrained to be $\ell=\pm1$. Without loss of generality one can restrict to $z_1>0$. The warp factor is fixed to be
\be
\me^{4 A}=\frac{z\big(3 p(z)-z p'(z)(1-k^3)\big)}{-p'(z) k}\, .
\ee 
The metric has the correct signature provided
\be
z p(z)\geq 0\, ,\quad -p'(z)\geq 0\, , \quad 0\leq k\leq 1\, .
\ee
Since the solution is invariant under the simultaneous reflection $z\rightarrow -z\, ,\, z_0\rightarrow -z_0$ we can further restrict to $z\geq 0$. The dilaton is given by
\be
\me^{4 \Phi}=\frac{1}{F_0^4}\frac{\big(3 p(z)-z p'(z)(1-k^3)\big)^3}{-p'(z)(3 p(z)-z p'(z))^2z^3 k^5}\, ,
\ee
while the fluxes may be succinctly written in terms of the potentials
\begin{align}
B&=-\frac{2}{3} \frac{z^2 p'(z)}{3 p(z)-z p'(z)}\dd k\wedge D\psi-\frac{k}{9}\frac{p'(z)-z p''(z)}{z}\vol({\Sigma_g})\, ,\\
C_1&=\frac{2F_0}{3}\frac{k z^2 p'(z)(1-k^3)}{3p(z)-z p'(z)(1-k^3)}D\psi\, ,\\
C_3&= \frac{2 F_0}{9} k^2\bigg[ \frac{p'(z)-z p''(z)}{3p(z)-z p'(z)(1-k^3)}p(z)+\frac{z p''(z)}{6}\bigg]D\psi \wedge \vol({\Sigma_g})\, ,
\end{align}
where
\begin{align}
H=\dd B\, ,\quad F_2=\dd C_1+F_0 B\, ,\quad F_4=\dd C_3+B\wedge F_2-\frac{1}{2} F_0 B\wedge B\, .
\end{align}
The above local solution has many different global completions depending on how the space is ended, see \cite{Bah:2017wxp}. Different degenerations of the space lead to the inclusion of different brane sources and thus different physics. We will review some of the possible degenerations briefly but refer to \cite{Bah:2017wxp} for further details. All the solutions we will present here allow for any of the global completions studied in \cite{Bah:2017wxp} however given the plethora of solutions we will focus on a single example for exposition.

There are various points where either the metric is singular or the circle parametrised by the coordinate $\psi$ shrinks. First consider where the $S^1$ shrinks at either $p(z)=0$ or $k=1$. For $z_1$ a single root of $p(z)$ the circle shrinks smoothly provided $\psi$ has period $2\pi$. Similarly at $k=1$ the circle shrinks smoothly if $\psi$ has period $2\pi$. Despite the two limits being separately smooth, the double limit is singular and corresponds to the presence of D6-branes with worldvolume AdS$_5\times \Sigma_g$.
The metric is singular at the three points $k=0$, $z=0$ and $p'(z)=0$. At $k=0$ and away from the two other degenerations the metric degenerates due to the presence of a stack of D8-branes on top of an O8-plane. There are smeared D4-branes located at $z=0$ and $p'(z)=0$. For the former the D4-branes are smeared along the Riemann surface, while for the latter the D4-branes are smeared along both the Riemann surface and the $S^2$. Note that $p'(z_1)=0$ provided $\kappa=\ell\neq 0$.
 
We may now fix the range of the coordinates. From the above we see that the coordinate ranges of $\psi$ and $k$ are uniquely fixed: $\psi$ has $2\pi$ period while $k\in[0,1]$. For the $z$ coordinate there are a larger number of options to take. 

For $\kappa=0$ or $\kappa=-\ell$, $p(z)$ only admits one real root at $z_0$, moreover $p'(z)$ has no real roots. Positivity of $p'(z)$ implies $\kappa=-\ell=-1$ and the $z$ coordinate is fixed between $z\in[0,z_0]$. There are smeared D4-branes at $z=0$ and a shrinking circle at $z=z_0$. 

Instead for $\kappa=\ell$, $p'(z)$ has roots at $z=\pm z_1$ and $p(z)$ can have three real roots. Writing 
\be
p(z)=\kappa(z-z_0)(z-z_-)(z-z_+)\, ,
\ee
where the roots satisfy
\be
z_0+z_-+z_+=0\, ,\quad 3 z_1^2=-z_0 z_--z_- z_+-z_+ z_0\, . 
\ee
We take $z_0$ to be real without loss of generality and then the other two roots are real if $z_0^2 \leq 4 z_1^2$. When the roots $z_{\pm}$ are complex the positivity conditions for the metric to be well-defined require
\be
z\in \begin{cases}
[0,z_1]&\text{for}\quad \kappa =1\, ,\, z_0<-2 z_1\, ,\\
[z_1,z_0]& \text{for}\quad \kappa=-1\, ,\, z_0> 2 z_1\, .
\end{cases}
\ee
For all three roots being real one finds that at least one root is always negative and at least one is always positive. The ranges are then
\be
z\in \begin{cases}
[0,z_0] & \text{for}\quad   \kappa=1\, ,\, 0<z_0\leq z_1\, ,\\
[z_1,z_0]& \text{for} \quad \kappa=-1\, ,\, z_1<z_0\leq 2 z_1\, .
\end{cases}
\ee

Having given the broad outline of the solutions we are able to present the truncation on this family of solutions. Following the truncation ansatz derived in section \ref{eq:theembedding} the metric is
\begin{align}
\dd s^2_{10}=&\me^{2 A}\bigg[g^{(5)}_{\mu\nu}dx^{\mu}dx^{\nu}- \frac{p'(z)}{9 z^2}\bigg( \dd s^2(\Sigma_g)+\frac{3 z\dd z^2}{p(z)} \nonumber\\
&+\frac{9 z^3}{3p(z)-z p'(z)}\bigg\{\frac{k}{1-k^3}\dd k^2 +\frac{4 (1-k^3)p(z)}{3\big(3 p(z)-z p'(z)(1-k^3)\big)}(D\psi-\mathcal{A})^2\bigg\}\bigg)\bigg]\, .
\end{align}
To compute the modification of the fluxes we must put them into the form used in \eqref{eq:Hform} and \eqref{eq:F0}-\eqref{eq:F4}. First we should identify the 1-form $\tilde{\xi}$ given in \eqref{eq:Hform}, 
\be
\tilde{\xi}=k \dd z+\frac{z\big(3 p(z)+z p'(z)\big)}{3 p(z)-z p'(z)}\dd k\, .
\ee
From the replacement rule in \eqref{eq:embeddign} it follows that the NSNS 3-form is
\be
H= \dd \Big(-\frac{k}{9}\frac{p'(z)-z p''(z)}{z}\vol({\Sigma_g})\Big)+\frac{2}{3} \frac{z^2 p'(z)}{3 p(z)-z p'(z)}\dd k\wedge \vol({\Sigma_g}) +\frac{1}{3} \Big({\cal D}\psi \wedge \dd \tilde{\xi}+\tilde{\xi}\wedge \mathcal{F}\Big)\, ,\label{eq:Hexgauged}
\ee
where ${\cal D}\psi = \dd \psi-A_g-\mathcal{A}$. The potential may be written as
\be
B=\frac{1}{3} \Big(\tilde{\xi}\wedge {\cal D}\psi +\frac{k(3 z^2 \kappa-p'(z))}{3 z}\vol({\Sigma_g})\Big)\, ,\label{eq:Bexgauge}
\ee
where the gauging is done through ${\cal D}\psi$ term and reproduces \eqref{eq:Hexgauged}.
The 2-form $F_2$ becomes
\be
F_2=\frac{1}{3} F_0\bigg(\tilde{\xi}\wedge {\cal D}\psi+\frac{ k (3 z^2 \kappa-p'(z))}{3 z}\vol({\Sigma_g}) \bigg) - \dd \bigg(\frac{F_0 k z\big(3 p(z)+(1-k^3)z p'(z)\big)}{3 \big(3 p(z)-(1-k^3) z p'(z)\big)}{\cal D}\psi\bigg)\, ,
\ee
where we may identify the first bracketed part as $F_0 B$ using the gauge given in \eqref{eq:Bexgauge}. Finally the 4-form flux is given by
\begin{align}
F_4=&\frac{1}{3 } F_2\wedge \tilde{\xi}\wedge {\cal D}\psi +\frac{F_0 k z}{18 }\dd (k z)\wedge {\cal D}\psi \wedge \vol({\Sigma_g})-\frac{F_0}{36}\dd\Big(k^2 p'(z) {\cal D}\psi\Big)\wedge \vol({\Sigma_g})+\\
&+\frac{F_0 k z}{3}\dd (k z)\wedge\Big( \star_5 \mathcal{F}-\frac{1}{3}\mathcal{F}\wedge {\cal D}\psi\Big)- F_0 \big(3 z^2 +4 p'(z)\big) \big(k^2 z \dd k +(k^3-2)\dd z\big) \wedge \vol({\Sigma_g})\wedge {\cal D}\psi\, .\nonumber
\end{align}

\subsection{Solutions of $d=5$ minimal gauged supergravity}

Let us now give some explicit supersymmetric solutions to $d=5$ minimal supergravity which may be uplifted to massive Type IIA supergravity using the results of the previous section. 

The first example that we will consider is a local solution, which possesses a number of different global completions. Our focus will be on two different global completions of this local solution. The first is a spindle while the second is constant curvature hyperbolic space. The local solution giving rise to both of these solutions is\footnote{This solution was first considered as a spindle in \cite{Boido:2021szx} in U$(1)^3$ gauged supergravity with the solution here obtained by taking the minimal limit ($A^1=A^2=A^3=\mathcal{A}$). The resultant solution is then a coordinate transformation (and rescaling owing to different conventions for the action) away from the solution in \cite{Ferrero:2020laf}. }
\begin{align}
g^{(5)}_{\mu\nu}dx^{\mu}dx^{\nu}&=P(y)^{1/3}\bigg( \dd s^2(\text{AdS}_3)+\frac{1}{4 q(y)}\dd y^2+ \frac{q(y)}{P(y)}\dd \phi^2\bigg)\, ,\nonumber\\
\mathcal{A}&=\frac{3 y}{y-a}\dd\phi\, ,\qquad P(y)=(y-a)^3\, ,\quad q(y)=P(y)-y^2\, .
\end{align}

\paragraph{Spindle solution}
Let us first consider the spindle, we will try to be as brief as possible since many of the details are by now well studied. For the space to be compact we require that the polynomial $q(y)$ has three real roots. It follows that this requires 
\be
-\frac{4}{27}\leq a\leq 0\, .
\ee
Note that the end-points of the interval are special. For $a=0$ the space is actually AdS$_5$ written with an AdS$_3$ slicing, while for $a=-\tfrac{4}{27}$ $q(y)$ has a double root at $y=\tfrac{8}{27}$. We will come back to this latter special point later. 

We should then fix $a$ to be strictly within this domain which implies that there are three single roots. It follows that two roots are necessarily positive while the third is necessarily negative. Let us denote the roots as $y_-, y_+, y_*$, with $y_-<0<y_+<y_*$. Then we must bound the $y$ coordinate as $y\in [y_-,y_+]$. At either end-point the space develops a conical deficit angle $2\pi (1- n_{\pm}^{-1})$ giving rise to the orbifold $\Sigma=\mathbb{WCP}^{1}_{[n_-,n_+]}$. The Euler characteristic of the space and magnetic charge of the solution are
\be
\chi(\Sigma)=\frac{1}{n_+}+\frac{1}{n_-}\, ,\qquad Q=\frac{1}{2\pi}\int_{\Sigma}\mathcal{F}=\frac{1}{n_-}-\frac{1}{n_+}\, ,
\ee
and thus exhibits an anti-twist, see \cite{Ferrero:2021etw}. Given the form of the roots we must take $n_+>n_->0$ and additionally require them to be relatively prime. In terms of the orbifold weights the period $\Delta \phi$, parameter $a$ and roots take the form:\footnote{To work this out it is simplest to solve for the roots $y_{\pm}$ and $a$ in terms of the third root and then to solve the period constraint.}
\begin{align}
\frac{\Delta \phi}{2\pi}&=\frac{n_+^2+ n_+n_-+n_-^2}{3 n_+n_- (n_++n_-)}\, ,\quad &a=&-\frac{ (n_+-n_-)^2 (2 n_++n_-)^2(2 n_-+n_+)^2}{27 (n_+^2+2 n_+ n_-+n_-^2)^3}\, ,\\
y_+&= \frac{(n_+-n_-)^3(2 n_++n_-)^3}{27 (n_+^2+2 n_+ n_-+n_-^2)^3}\, ,&y_-=&-\frac{(n_+-n_-)^3(2 n_-+n_+)^3}{27 (n_+^2+2 n_+ n_-+n_-^2)^3}\, .
\end{align}

\paragraph{Hyperbolic space}
We noted earlier that $a=-\tfrac{4}{27}$ is a special point where the function $q(y)$ develops a non-trivial double root. As we will show, by taking a certain scaling limit to this point we obtain the metric on a constant curvature hyperbolic disc. To wit, set $a=-\tfrac{4}{27}$ and define
\be
y = \frac{8}{27}+\epsilon Y\, ,\qquad \phi= \frac{4\chi}{9\epsilon}\, .
\ee
Expanding the metric around $\epsilon=0$ we find
\be
g^{(5)}_{\mu\nu}dx^{\mu}dx^{\nu}=\frac{4}{9}\bigg[\dd s^2(\text{AdS}_3)+\frac{3}{4} \Big(\frac{\dd Y^2}{Y^2}+Y^2 \dd \chi^2\Big)\bigg]\, ,
\ee
which is the direct product of AdS$_3$ and $\mathbb{H}^2$. The gauge field works similarly, though one must add in a pure gauge term
\be
A\rightarrow A-\frac{8}{9 \epsilon}\dd \chi\overset{\epsilon\rightarrow 0}{\longrightarrow} Y \dd\chi\, .
\ee
The normalisation of the metric and gauge field implies
\be
2(1-g)=\chi(\mathbb{H}^2)=\frac{1}{4\pi}\int_{\mathbb{H}^2} R\,  \vol(\mathbb{H}^2)= -\frac{1}{2\pi} \text{Vol}(\mathbb{H}^2)=-\frac{1}{2\pi}\int_{\mathbb{H}^2}\mathcal{F}\, .
\ee
We therefore see that the magnetic charge perfectly cancels the Euler characteristic and supersymmetry is preserved via a topological twist.

\paragraph{Gutowski--Reall black hole}
The second class of solution that we may uplift is the Gutowski--Reall black hole solution \cite{Gutowski:2004ez}.
This is an asymptotically AdS$_5$ rotating black hole of $d=5$ minimal gauged supergravity,\footnote{To embed this into the class of solutions in section \ref{sec:uplift} one must fix $m=1$ below.}
 \begin{align}
g_{\mu\nu}^{(5)}dx^{\mu}dx^{\nu}&= -\frac{u}{\Lambda}dt^2+ \frac{dr^2}{u}+\frac{r^2}{4}\left(L_1^2+L_2^2+\Lambda(L_3-\Omega dt)^2\right),\nn\\[2mm]
{\cal A}&=-3\left(\left(1-\frac{r_0^2}{r^2}- \frac{m^2 r_0^4}{2r^2}\right)dt+\epsilon \frac{m r_0^4}{4 r^2}L_3\right),~~~~dL_i=\frac{1}{2}\epsilon_{ijk}L_j\wedge L_k,\nn\\[2mm]
u&=\left(1-\frac{r_0^2}{r^2}\right)^2\left(1+m^2(r^2+2 r_0^2)\right),~~~~\Lambda=1+m^2\left(\frac{r_0^6}{r^4}- \frac{r_0^8}{4 r^6}\right),\nn\\[2mm]
\Omega&=\frac{2 m \epsilon}{\Lambda}\bigg[\left(\frac{3}{2}+r_0^2m^2\right)\frac{r_0^4}{r^4}-\left(\frac{1}{2}+\frac{m^2 r_0^2}{4}\right)\frac{r_0^6}{r^6}\bigg],~~~~\epsilon^2=1.
\end{align}

We may now simply insert the solutions outlined here into the uplift worked out in the previous section \ref{sec:uplift} to obtain new solutions of massive Type IIA supergravity. 
It is interesting to understand the form of the solutions that we obtain. Recall that the seed solution on which we performed the truncation can be interpreted as the holographic duals of the IR limit of wrapping one of the six-dimensional, $\mathcal{N}=(1,0)$ theories studied in \cite{Gaiotto:2014lca} on a constant curvature Riemann surface. For the case of the solutions of $d=5$ minimal gauged supergravity on a spindle and hyperbolic space we may interpret the uplifted solutions as the holographic duals of two-dimensional SCFTs obtained by compactifying the six-dimensional, $\mathcal{N}=(1,0)$ theory on the four-manifold consisting of the direct product of the seed Riemann surface with either a spindle or two-dimensional hyperbolic space, see for example \cite{Faedo:2021nub,Giri:2021xta,Cheung:2022ilc,Suh:2022olh} for similar setups in different theories. 
A similar interpretation for the uplift of the Gutowski--Reall solution is somewhat more subtle. It would be interesting to understand the thermodynamics of the black hole in this uplift and to identify the microstates of the black hole.

\section*{Acknowledgments}
We thank Emanuel Malek for clarifying correspondence. CC is supported by the National Research Foundation of Korea (NRF) grant numbers 2019R1A2C2004880 and 2020R1A2C1008497.
NM  is supported by AEI-Spain (under project PID2020-114157GB-I00 and Unidad de Excelencia Mar\'\i a de Maetzu MDM-2016-0692), by Xunta de Galicia-Conseller\'\i a de Educaci\'on (Centro singular de investigaci\'on de Galicia accreditation 2019-2022, and project ED431C-2021/14), and by the European Union FEDER. The work of AP was funded, in whole, by ANR (Agence Nationale de la Recherche) of France under contract number ANR-22-ERCS-0009-01. 
AP would like to thank the National and Kapodistrian University of Athens for hospitality during completion of this work.
\appendix

\section{Conventions}\label{sec:appendix}
In this appendix we spell out our conventions. First off, we shall use Roman and Greek letters (the latter reserved for the external space) to indicate curved indices and underline them to indicate flat indices.
We follow the conventions of \cite{Tomasiello:2011eb} with the hodge dual defined as
\beq
\star \e^{\underline{M}_1...\underline{M}_k}=  \frac{1}{(d-k)!}\epsilon_{\underline{M}_{k+1}...\underline{M}_{d-k}}^{~~~~~~~~~~~\underline{M}_1...\underline{M}_k} \e^{\underline{M}_{k+1}...\underline{M}_{d-k}},
\eeq
The self-duality relation for the $d=10$ polyform flux is 
\beq
F =  \star \lambda(F),\label{eq:selfduality}
\eeq
where $\lambda(C_k)= (-1)^{\lfloor\frac{k}{2}\rfloor} C_k$, for a $k$-form $C_k$. We define the matrix
\beq
\hat \gamma = \eta \gamma^{1....d}= \eta (-1)^t \gamma_{1....d}
\eeq
in all dimensions and signatures, where $t$ is the number of time-like directions, such that
\beq
\eta^2=  (-1)^t (-1)^{[\frac{d}{2}]}.
\eeq
In odd dimensions we shall fix
\beq
\hat\gamma = \mathbb{I}
\eeq
while in even dimensions $\hat\gamma$ is the chirality matrix. We assume, as \cite{Tomasiello:2011eb} does, that in ten dimensions
\beq
\eta^{(10)}=1. 
\eeq
Through the Clifford map we have the following action on forms
\beq
\hat \gamma C_k= \eta \star \lambda(C_k),~~~~ C_k \hat\gamma = \eta (-)^{[\frac{d}{2}]} \lambda(\star C_k)\label{eq:hodualtogamma}.
\eeq
In $d=10$ we also define
\beq
\overline{\epsilon}=  (\Gamma_0 \epsilon)^{\dag}= \epsilon^{\dag}\Gamma^0,
\eeq
and make use of some shorthand notation
\begin{align}
C_{M}&= \iota_{dx^M} C,~~~~~ C^2= \sum_{k} \frac{1}{k!} (C_k)_{M_1...M_k}(C_k)^{M_1...M_k},~~~~\nonumber\\
 C^2_{MN}&=\sum_{k} \frac{1}{(k-1)!} (C_k)_{MM_1...M_{k-1}}(C_k)_N^{~~M_1...M_{k-1}}.
\end{align}
Note also
\beq
\sqrt{|g|}\frac{1}{k!} (C_k)_{M_1...M_k}(C_k)^{M_1...M_k}= \star C_k\wedge C_k.
\eeq
We shall be interested in a split of the gamma matrices into $10=5+5$, as such we shall parameterise them as
\beq
\Gamma_{\underline{\mu}}= \sigma_3\otimes\gamma_{\underline{\mu}}\otimes \mathbb{I},~~~~\Gamma_{\underline{a}}= \sigma_1\otimes \mathbb{I}\otimes \gamma_{\underline{a}},
\eeq
where we have split the $d=10$ index $M=(\mu,a)$ for $\mu$ an index on a $d=5$ Lorentzian space and $a$ an index on a $d=5$ Euclidean space.
We work in conventions where  $\gamma_{12345}=1$, as such
\beq
\hat\Gamma =\sigma_2\otimes \mathbb{I}\otimes \mathbb{I}.
\eeq
The intertwiner defining $d=10$ Majorana conjugation (m.c.)  as $\epsilon^c=B^{(10)}\epsilon^{*}$ then decomposes in terms of correspoding intertwiners $\tilde{B}$ on the external and $B$ on the internal space as
\beq
B^{(10)}=\sigma_1\otimes \tilde{B}\otimes B\, ,
\eeq
where
\beq
\tilde{B}^{-1}\gamma_{\mu}\tilde{B}=-\gamma_{\mu}^{*},~~~~ \tilde{B} \tilde{B}^*= -\mathbb{I},~~~~~ \tilde{B}^{\dag}=\tilde{B} \label{eq:extB}.
\eeq
and
\beq
B^{-1}\gamma_{a}B=\gamma_{a}^{*},~~~~ BB^*= -\mathbb{I},~~~~~ B^{\dag}=B,
\eeq
Finally we define the spin covariant derivative as
\beq
\nabla_M= \partial_{M}+ \frac{1}{4}\omega_M{}^{\underline{P}\underline{Q}}\Gamma_{\underline{P}\underline{Q}},~~~~ d\e^{\underline{M}}+\omega^{\underline{M}}{}_{\underline{N}}\wedge \e^{\underline{N}}=0,
\eeq
and spinorial Lie derivative  as
\beq
{\cal L}_{K}\epsilon= K^M\nabla_{M}\epsilon+ \frac{1}{4}\nabla_{M}K_{N}\Gamma^{MN}\epsilon.
\eeq

\section{Some details of $d=5$ Lorentzian bi-linears}\label{eq:deq5lordetails}
In this appendix we provide some details of the $d=5$ Lorentzian bi-linears we refer to in the main text.\\
~~\\
In terms of a generic Dirac spinor in $d=5$ Lorentzian space, $\zeta$, we define the following bi-spinors
\beq
\phi^1= \zeta \otimes \overline{\zeta},~~~~\phi^2= \zeta \otimes \overline{\zeta^c},
\eeq
for which (given \eqref{eq:extB}) all of $\phi^1$ but only $(\phi^2)_{2,3}$ are non-trivial and
\beq
\phi^{1,2}=i \star_5 \lambda(\phi^{1,2}).
\eeq
This suggests defining 
\beq
i f \equiv \overline{\zeta}\zeta,~~~~ k_{\mu} \equiv \overline{\zeta}\gamma_{\mu}\zeta,~~~~X_{\mu\nu} \equiv \overline{\zeta}\gamma_{\mu \nu}\zeta,~~~~Y_{\mu\nu} \equiv \overline{\zeta^c}\gamma_{\mu \nu}\zeta,
\eeq
so that
\beq
\phi^1= \frac{1}{4}(1+ i \star_5 \lambda)(i f+k-X)=\frac{f}{4}\left(i+\frac{k}{f}\right)\wedge e^{i\frac{X}{f}},~~~~\phi^2= -\frac{1}{4}(1+ i \star_5 \lambda) Y=\frac{f}{4}\left(1-i \frac{k}{f}\right)\wedge \frac{Y}{f},
\eeq
where $(f,k,X)$ are real and $Y$ complex and we do not attempt to refine things further as  $f$ is not necessarily non-vanishing.
Note also that
\beq\label{eq:5didentities}
k\zeta= i f \zeta,~~~~X\zeta=-2 i f \zeta,~~~~~ \iota_k k =- f^2,~~~~\iota_k\phi^{1,2}_-=  i f \phi^{1,2}_+,~~~ k\wedge \phi^{1,2}_+= i f \phi^{1,2}_-
\eeq
and that
\beq
Y\wedge X=0,~~~~~X\wedge X=\frac{1}{2}Y\wedge \overline{Y},~~~~\iota_{k}X=\iota_k Y=0.
\eeq
Note that we have not used \eqref{eq:externalSUSY} to derive any of these conditions, they are completely general.

\section{Proving sufficiency of the embedding}\label{sec:suff}
In appendix \ref{eq:susysuff} we prove that the embedding of $d=5$ minimal supergravity, presented in the main text, indeed preserves $d=10$ supersymmetry, when the background on the external space preserves $d=5$ supersymmetry. Furthermore, in appendix \ref{eq:EOMsuff}, we prove that the embedding gives a solution to the $d=10$ equations of motion, even when the solution of the $d=5$ supergravity is not supersymmetric. For clarity's sake, let us stress that through out this appendix we are using the term AdS$_5$ vacua loosely. We include also the limit where the inverse radius $m=0$, i.e. the Mink$_5$ vacua limit, where it is ungauged supergravity that is being embedded in ten dimensions. 

\subsection{Sufficiency for supersymmetry}\label{eq:susysuff}
In this appendix we prove that our embedding of $d=5$ minimal gauged supergravity into massive Type IIA supergravity preserves supersymmetry in $d=10$ provided that it preserves supersymmetry in $d=5$. We find it easiest to do this in terms of the necessary spinorial conditions.\\
~~\\
A solution of Type IIA supergravity preserves supersymmetry if it supports two Majorana--Weyl Killing spinors such that the gravitino and dilatino variations, respectively
\begin{align}
\delta\psi^1_{M}&=\Big(\nabla^{(10)}_M-\frac{1}{4}H_M\Big)\epsilon_1+ \frac{e^{\Phi}}{16}F\Gamma_{M}\epsilon_2,\nn\\[2mm]
\delta\psi^2_{M}&=\Big(\nabla^{(10)}_M+\frac{1}{4}H_M\Big)\epsilon_2+ \frac{e^{\Phi}}{16}\lambda(F)\Gamma_{M}\epsilon_1,\nn\\[2mm]
\delta \lambda^1&=\Big(-\frac{1}{2}H+ d\Phi\Big)\epsilon_1+\frac{e^{\Phi}}{16}\Gamma^MF\Gamma_M\epsilon_2,\nn\\[2mm]
\delta \lambda^2&=\Big(\frac{1}{2}H+ d\Phi\Big)\epsilon_2+\frac{e^{\Phi}}{16}\Gamma^M\lambda(F)\Gamma_M\epsilon_1 \label{eq:susyvariations},
\end{align}
all vanish. For the case at hand the fields  $(F,H,\Phi)$ and metric are defined in \eqref{eq:embeddign}, the flat space gamma matrices in the preceding appendix and the Killing spinors are as in \eqref{eq:spinoransatz} with $\zeta$ obeying \eqref{eq:externalSUSY}.

Let us begin by considering the vanishing of the gravitino variations, which requires us to decompose the covariant derivative and curved space gamma matrices on a space of the form
\beq
ds^2_{10}= e^{2A} g^{(5)}_{\mu\nu} dx^{\mu}dx^{\mu}+ ds^2(\text{M}_4)+ e^{2C}{\cal D}\psi^2,~~~~ {\cal D}\psi\equiv d\psi+ V-{\cal A}.
\eeq
One can show that the spin covariant derivative on this space decomposes as
\begin{align}
\nabla^{(10)}_{\mu}&= \nabla_{\mu}-{\cal A}_{\mu}(\nabla_{\psi}-\partial_{\psi})+\frac{1}{2}(\Gamma_{\mu}+{\cal A}_{\mu}\Gamma_{\psi})\partial A+\frac{1}{4}\Gamma_{\psi}{\cal F}_{\mu},\nn\\[2mm]
\nabla^{(10)}_i&=\nabla_{i}+ V_i(\nabla_{\psi}-\partial_{\psi})- \frac{1}{4}\Gamma_{\psi} (dV)_i ,\nn\\[2mm]
\nabla^{(10)}_\psi&=\nabla_{\psi}+\frac{1}{4}e^{2C}{\cal F},~~~~\nabla_{\psi}= \partial_{\psi}+\frac{1}{2}\Gamma_{\psi} \partial C-\frac{e^{2C}}{4} dV
\end{align}
where we have further split the internal index as $a=(\psi,i)$. The gamma matrices likewise decompose as
\beq
\Gamma_{\mu}= e^{A} \sigma_3\otimes \gamma_{\mu}\otimes \mathbb{I}-{\cal A}_{\mu} \sigma_1\otimes\mathbb{I}\otimes \gamma_{\psi},~~~~\Gamma_i= \sigma_1\otimes \mathbb{I}\otimes (\gamma_i+V_i \gamma_{\psi}),~~~~\gamma_{\psi}=e^C\gamma_{\underline{\psi}}.
\eeq
To proceed we observe that \eqref{eq:selfduality} and \eqref{eq:hodualtogamma} together imply that
\beq
\hat\Gamma F= F,~~~ \hat \Gamma \lambda (F)=- \lambda (F),
\eeq
allowing us to simplify the RR flux terms in the gravitino variations a little, for instance
\beq
F\Gamma_{M}\epsilon_2=( 1+\hat \Gamma)\Big(f_+-\frac{4}{3 c} e^{-\Phi} {\cal F}\wedge \text{Im}\psi^1_+\Big)\Gamma_{M}\epsilon_2= 2\Big(f_+-\frac{4}{3 c} e^{-\Phi} {\cal F}\wedge \text{Im}\psi^1_+\Big)\Gamma_{M}\epsilon_2.
\eeq
We also find it helpful to bring the external gravitino condition to the form
\beq
\Big(\nabla_{\mu}+\frac{i}{2}m {\cal A}_{\mu}\Big)\zeta= \Big(\frac{m}{2}\gamma_{\mu}\zeta-\frac{3i}{24}{\cal F}\gamma_{\mu}+\frac{i}{24}\gamma_{\mu}{\cal F}\Big)\zeta,\label{eq:extennalkillinspinorequation}
\eeq
and write the $d=10$ spinors as
\beq
\epsilon_1= \theta_+\otimes \bigg[\zeta\otimes \chi_1-i \zeta^c\otimes \chi_1^c\bigg],~~~~\epsilon_2= \theta_-\otimes \bigg[\zeta\otimes \chi_2+i \zeta^c\otimes \chi_2^c\bigg],
\eeq
with $\theta_{\pm}$ short hand for the auxiliary vectors appearing in \eqref{eq:spinoransatz}. 
To make progress the important thing to appreciate is that when we fix ${\cal A}=0$, the supersymmetry variations \eqref{eq:susyvariations} reduce to those of AdS$_5$ vacua once the external spinors have been factored out, so
\beq
\delta\psi^1_{M}\bigg\lvert_{{\cal A}=0}=\delta\psi^2_{M}\bigg\lvert_{{\cal A}=0}=0,
\eeq
or equivalently
\begin{align}
(m e^{-A}+ i \partial A)\chi_1 + \frac{e^{\Phi}}{4} f_+ \chi_2=0,~~~~(m e^{-A}- i \partial A)\chi_2 + \frac{e^{\Phi}}{4} \lambda(f_+) \chi_1&=0,\nn\\[2mm]
\Big(\nabla_{\psi}-\frac{1}{4}H_2\Big)\chi_1-i\frac{e^{\Phi}}{8}f_{+}\gamma_{\psi}\chi_2=0,~~~~\Big(\nabla_{\psi}+\frac{1}{4}H_2\Big)\chi_2+i\frac{e^{\Phi}}{8}\lambda (f_{+})\gamma_{\psi}\chi_1&=0,\nn\\[2mm]
\Big(\nabla_i-V_i \partial_{\psi}-\frac{1}{4}\gamma_{\psi}\big(V_i-(H_2)_i\big)-\frac{1}{4}(H_3)_i\Big)\chi_1- i\frac{e^{\Phi}}{8}f_+\gamma_i\chi_2&=0,\nn\\[2mm]
\Big(\nabla_i-V_i \partial_{\psi}-\frac{1}{4}\gamma_{\psi}\big(V_i+(H_2)_i\big)+\frac{1}{4}(H_3)_i\Big)\chi_2- i\frac{e^{\Phi}}{8}\lambda(f_+)\gamma_i\chi_1&=0,
\end{align}
hold by definition as the internal space (modulo the ${\cal A}$ dependence in the fibre) is that of the AdS$_5$ vacua. Our task then is to show that $\delta\psi^{1,2}_{M}=\delta\psi^{1,2}_{M}\bigg\lvert_{{\cal A}=0}$ when \eqref{eq:extennalkillinspinorequation} is assumed to hold for a non-trivial $\zeta$. To this end the following identities are useful
\beq
\partial_{\psi}\chi_{1,2}= \frac{i}{2}m \chi_{1,2},~~~~\gamma_{\psi}= \frac{e^{C}}{||\xi||}\xi,~~~~\text{Im}\lambda (\psi^1_+)\xi\chi_1= \frac{i}{2}||\xi||^2\chi_2,~~~~\text{Im}\psi^1_+\xi\chi_2= \frac{i}{2}||\xi||^2\chi_1.
\eeq
These and the rest of the identities we use can be easily proved with a concrete representative spinor $\chi$ that gives rise to the internal bi-spinors $\psi^{1,2}$ as in section \ref{sec:onAdS5vaccua}. Using these it is now easy to show that 
\begin{align}
&\delta\psi^{1}_{\psi}-\left(\delta\psi^{1}_{\psi}\bigg\lvert_{{\cal A}=0}\right)= \frac{e^{C-2A}}{4}\bigg[e^C-\frac{||\xi||}{3c}\bigg]\theta_{+}\otimes{\cal F} \zeta \otimes \chi_{1}+\text{m.c.},\nn\\[2mm]
&\delta\psi^{2}_{\psi}-\left(\delta\psi^{2}_{\psi}\bigg\lvert_{{\cal A}=0}\right)= \frac{e^{C-2A}}{4}\bigg[e^C-\frac{||\xi||}{3c}\bigg]\theta_{-}\otimes{\cal F} \zeta \otimes \chi_{2}+\text{m.c.},
\end{align}
which is zero on our classes of solutions by definition. Likewise, using the identities 
\beq
\text{Im}\psi^1_+ \chi_2=-\frac{i}{2} e^A c \chi_1,~~~~\text{Im}\lambda(\psi^1_+) \chi_1=\frac{i}{2} e^A c \chi_2,~~~~\xi \chi_1=e^Ac( \chi_1- \overline{a}\chi_2),~~~~\tilde{\xi} \chi_1= e^A c\overline{a}\chi_2,
\eeq
we  find
\begin{align}
\delta\psi^{1}_{\mu}+{\cal A}_{\mu}\delta\psi^{1}_{\psi}-\left(\delta\psi^{1}_{\mu}\bigg\lvert_{{\cal A}=0}\right)&={\cal A}_{\mu}\theta_{+}\otimes \zeta \otimes \bigg[\partial_{\psi}\chi_{1}-\frac{i}{2}m \chi_{1}\bigg]\nn\\[2mm]
&+\frac{i e^{-A}}{4 ||\xi||}\bigg[e^{C}-\frac{||\xi||}{3c}\bigg]\theta_{+}\otimes{ \cal F}_{\mu} \zeta\otimes(\tilde\xi\chi_{1}- e^A c\chi_{1})+ \text{m.c.},\nn\\[2mm]
\delta\psi^{2}_{\mu}+{\cal A}_{\mu}\delta\psi^{2}_{\psi}-\left(\delta\psi^{2}_{\mu}\bigg\lvert_{{\cal A}=0}\right)&={\cal A}_{\mu}\theta_{-}\otimes \zeta \otimes \bigg[\partial_{\psi}\chi_{2}-\frac{i}{2}m \chi_{2}\bigg]\nn\\[2mm]
&+\frac{i e^{-A}}{4 ||\xi||}\bigg[e^{C}-\frac{||\xi||}{3c}\bigg]\theta_{-}\otimes{ \cal F}_{\mu} \zeta\otimes(\tilde\xi\chi_{2}- e^A c\chi_{2})+ \text{m.c.},
\end{align}
where every term in square brackets is necessarily zero. Finally, one can show that
\begin{align}
&\delta\psi^{1}_{i}-V_i\delta\psi^{1}_{\psi}-\left(\delta\psi^{1}_{i}\bigg\lvert_{{\cal A}=0}\right)= \frac{e^{-2A}}{12c}\theta_{+}\otimes {\cal F}\zeta\otimes \bigg[- \tilde{\xi}_i\chi_{1}+2i \text{Im}\psi^1_+\gamma_i\chi_{2}\bigg]+\text{m.c.},\nn\\[2mm]
&\delta\psi^{2}_{i}-V_i\delta\psi^{2}_{\psi}-\left(\delta\psi^{2}_{i}\bigg\lvert_{{\cal A}=0}\right)= \frac{e^{-2A}}{12c}\theta_{-}\otimes {\cal F}\zeta\otimes \bigg[- \tilde{\xi}_i\chi_{2}+2i \text{Im}\lambda(\psi^1_+)\gamma_i\chi_{1}\bigg]+\text{m.c.},
\end{align}
where the right-hand side vanishes via the identities
\beq
\text{Im}\psi^1_+\gamma_i\chi_2=-\frac{i}{2}\tilde{\xi}_i\chi_1,~~~~\text{Im}\lambda(\psi^1_+)\gamma_i\chi_1=\frac{i}{2}\tilde{\xi}_i\chi_2.
\eeq
This exhausts all the directions of the gravitino variations. Moving now onto the dilatino variations, we again have by definition that
\beq
\delta\lambda^{1}\bigg\lvert_{{\cal A}=0}=\delta\lambda^{2}\bigg\lvert_{{\cal A}=0}=0
\eeq
holds because the internal space is fixed so as to match that of the AdS$_5$ vacua. It is simple to show that
\begin{align}
\delta\lambda^{1}-\left(\delta\lambda^{1}\bigg\lvert_{{\cal A}=0}\right)&=\frac{e^{-2A}}{6c}\theta_{-}\otimes {\cal F}\zeta\otimes \bigg[i\tilde{\xi}\chi_{1}-(\gamma^a\text{Im}\psi^1_+\gamma_a+\text{Im}\psi^1_+)\chi_2\bigg]+\text{m.c.},\\[2mm]
\delta\lambda^{2}-\left(\delta\lambda^{2}\bigg\lvert_{{\cal A}=0}\right)&=\frac{e^{-2A}}{6c}\theta_{+}\otimes {\cal F}\zeta\otimes \bigg[i\tilde{\xi}\chi_{2}+(\gamma^a\text{Im}\lambda(\psi^1_+)\gamma_a+\text{Im}\lambda(\psi^1_+))\chi_1\bigg]+\text{m.c.},
\end{align}
which vanish if the quantities in square brackets sum to zero, which indeed turns out to be the case for the bi-linears and spinors defined as in section \ref{sec:onAdS5vaccua}. This completes our proof that the embedding of any supersymmetric solution of $d=5$ minimal supergravity into massive Type IIA supergravity preserves $d=10$ supersymmetry.

\subsection{Sufficiency for equations of motion}\label{eq:EOMsuff}
In this appendix we prove that the embedding of any solution of $d=5$ minimal supergravity gives rise to a solution of the Type IIA supergravity equations of motion, irrespective of whether external supersymmetry is assumed to hold or not.\\
~~\\
The Type IIA supergravity equations of motion and Bianchi identities, away from the loci of any possible sources, take the form
\begin{align}
&d_H F=0,~~~~ dH=0,~~~~{\cal H}\equiv d(e^{-2\Phi}\star_{10}H)-\frac{1}{2}(F,F)_8=0,
\nn\\[2mm]
&\mathcal{D}\equiv 2R^{(10)}- H^2-8 e^{\Phi}(\nabla^{(10)})^2 e^{-\Phi}=0,~~~ {\cal E}_{AB}\equiv R^{(10)}_{AB}+2 \nabla^{(10)}_{A}\nabla^{(10)}_{B}\Phi-\frac{1}{2} H^2_{AB}-\frac{e^{\Phi}}{4} (F)^2_{AB}=0.
\end{align}
We already establish that the first two of these hold in the main text, as 
\beq
dH=dH\bigg\lvert_{{\cal A}=0}=0,~~~~d_H F=d_H F\bigg\lvert_{{\cal A}=0}=0,
\eeq
with the first equality following from the Bianchi identity and equation of motion of the ${\cal F}$ and the second because what is left is equal to the AdS$_5$ vacua result, which we know vanishes. For the equation of motion of the NSNS flux it should be clear from the form of $(F,H)$ (see \eqref{eq:embeddign}) that it decomposes into parts parallel to $\text{vol}_5$ which are  implied because they hold for AdS$_5$ vacua so we will not quote explicitly, and the parts defined by the following 
\begin{align}
e^{-2\Phi}\star_{10}H&=  \frac{e^{A-2\Phi}}{3c} \star\tilde{\xi}\wedge \star_5{\cal F}+...,\\[2mm]
\frac{1}{2}(F,F)_8&= \frac{8 e^{-2\Phi}}{9 c^2}({\cal F}\wedge \text{Im}\psi^1_+,{\cal F}\wedge \text{Im}\psi^1_+)_8+ \frac{2 e^{A-\Phi}}{3c}\bigg[(f_+,\star_5{\cal F}\wedge\text{Im}\psi^1_-)_8+(\star_5{\cal F}\wedge\text{Im}\psi^1_-,f_+)_8\bigg]\nn\\[2mm]
&- \frac{2 e^{-\Phi}}{3c}\bigg[(f_+,{\cal F}\wedge\text{Im}\psi^1_+)_8+({\cal F}\wedge\text{Im}\psi^1_+,f_+)_8\bigg]\nn\\[2mm]
&-\frac{8 e^{A-2\Phi}}{9c^2}\bigg[({\cal F}\wedge \text{Im}\psi^1_+,\star_5{\cal F}\wedge\text{Im}\psi^1_-)_8+(\star_5{\cal F}\wedge\text{Im}\psi^1_-,{\cal F}\wedge \text{Im}\psi^1_+)_8\bigg]+...
\end{align}
To show this is implied some identities are necessary; first off with the properties of the pairing  one can establish that
\begin{align}
&({\cal F}\wedge \text{Im}\psi^1_+,{\cal F}\wedge \text{Im}\psi^1_+)_8=  -\frac{1}{8}e^{A} c{\cal F}\wedge {\cal F} \wedge \star\tilde \xi,\\[2mm]
&(f_+,{\cal F}\wedge\text{Im}\psi^1_+)_8+({\cal F}\wedge\text{Im}\psi^1_+,f_+)_8=0,\nn\\[2mm]
&(f_+,\star_5{\cal F}\wedge\text{Im}\psi^1_-)_8+(\star_5{\cal F}\wedge\text{Im}\psi^1_-,f_+)_8= \star_5 {\cal F}\wedge( \text{Im}\psi^1_+, \star \lambda( f_+)+\frac{1}{c e^A} (\iota_{\xi}+\tilde{\xi})\wedge f_+)_5,\nn\\[2mm]
&= \frac{4}{c}\star_5 {\cal F}\wedge( \text{Im}\psi^1_+,e^{-5A}(d_H(e^{4A-\Phi}\text{Im}\psi^1_+)-4m e^{3A-\Phi}\text{Re}\psi^1_-)- e^{-A}d_H(e^{-\Phi} \text{Im}\psi^1_+))_5,\nn\\[2mm]
&= \frac{16}{c} e^{-A-\Phi}\star_5 {\cal F}\wedge( \text{Im}\psi^1_+,dA\wedge  \text{Im}\psi^1_+)_5= 2 e^{-\Phi}\star_5 {\cal F}\wedge dA\wedge \star \tilde{\xi},\\[2mm]
&({\cal F}\wedge \text{Im}\psi^1_+,\star_5{\cal F}\wedge\text{Im}\psi^1_-)_8+(\star_5{\cal F}\wedge\text{Im}\psi^1_-,{\cal F}\wedge \text{Im}\psi^1_+)_8=-\frac{1}{8}||\xi|| {\cal F}\wedge \star_5 {\cal F}\wedge \star_4 \tilde{\xi},
\end{align}
where we have used the non-trivial fact that
\beq
e^{4A}\bigg[d_H(e^{-\Phi} \text{Im}\psi^1_+)+\frac{1}{4}(\iota_{\xi}+\tilde{\xi}\wedge) f_+\bigg]-\bigg[d_H(e^{4A-\Phi}\text{Im}\psi^1_+)-4m e^{3A-\Phi}\text{Re}\psi^1_--\frac{c}{4}e^{5A}\star \lambda (f_+)\bigg]=0
\eeq
holds in general, the ${\cal A}$ dependence cancelling between the two terms and we also use that $(\text{Im}\psi^1_+,\text{Re}\psi^1_-)_5=0$.
To deal with $d\star \tilde{\xi}$ we use an identity that must hold given that a time-like Killing vector can be defined on AdS$_5$ (which leads to a time-like Killing vector in $d=10$)
\begin{align}
&d(e^{-2\Phi}\star_{10}\tilde K^{(10)})\bigg\lvert_{{\cal A}=0}=0 ~~~~\Rightarrow ~~~~  d (e^{5A-2\Phi} \star \tilde{\xi})\bigg\lvert_{{\cal A}=0}=0\\
&\Rightarrow ~~~~   d (e^{5A-2\Phi} \star \tilde{\xi}) =-e^{5A-2\Phi} \frac{||\xi||}{3c} {\cal F}\wedge \star_4\tilde{\xi}
\end{align}
(see \cite{Legramandi:2018qkr} where we have corrected an obvious typo, given the conditions that lead to (D.3) there). With this it is now possible to  establish that \eqref{eq:intsusy1ads5s}-\eqref{eq:intsusy7s} and the external flux equation of motion imply
\beq
\frac{1}{2}(F,F)= \frac{1}{3c}d(e^{A-2\Phi} \star\tilde{\xi}\wedge \star_5{\cal F})+....= d(e^{-2\Phi}\star H),
\eeq
so the NSNS flux equation of motion is implied.  We now pause to make an observation: in \cite{Legramandi:2018qkr} it is proven that supersymmetry plus the equations of motions and Bianchi identities of the fluxes imply the remaining equations of motion when $K^{(10)}$ is assumed to be time-like. We further note that it is possible to show that $K^{(10)}$ is time-like or null if and only if the external Killing vector $k^{\mu}\partial_{\mu}$ is like-wise time-like or null. However as \eqref{eq:embeddign} does not depend on any of the external bi-linears (including $k$), and for ${\cal A}=0$ we know all the $d=10$ equations of motion hold,  then in general the $d=10$ equations of motion must be closing on the equations of motion of $d=5$ minimal gauged supergravity. As such if these are implied for a time-like supersymmetric solution of this theory, they should be implied for all solutions of the theory. If one is willing to trust that argument, we have already shown what we set out to, however we are aware that the reader may be unsatisfied with this, for this reason we will now also complete the proof in a more direct way.\\
~~\\
To proceed we must solve Einsteins equations and the dilaton equation of motion. We will achieve this in a similar fashion to how we established that supersymmetry holds, i.e.\ by assuming that the equations of motion of $d=5$ minimal gauged supergravity \eqref{eq:exein}-\eqref{eq:exfluxeom} hold  we shall show that the equations of motion for generic ${\cal A}$ reduce to  those of the AdS$_5$ vacua for which ${\cal A}=0$. To this end we need to decompose several of the objects appearing here in terms of a metric of the form
\beq
ds^2_{10}= e^{2A} g^{(5)}_{\mu\nu}dx^{\mu}dx^{\nu}+ ds^2(\text{M}_4)+ e^{2C}{\cal D}\psi^2.
\eeq
Again decomposing the internal directions as $a=(\psi,i)$, the Ricci tensor and scalar decomposes in coordinate frame as
\begin{align}
R^{(10)}&=R^0+\frac{e^{-4A}}{18}(e^{2A}-9e^{2C}){\cal F}^2,~~~~R^{(10)}_{\psi\psi}= R^0_{\psi\psi}+\frac{e^{-4(A-C)}}{2}{\cal F}^2,~~~~R^{(10)}_{ij}= R^0_{ij}-V_i V_jR^{(10)}_{\psi\psi}+2 V_{(i}R^{(10)}_{j)\psi},\nn\\[2mm]
R^{(10)}_{\mu\psi}&= -{\cal A}_{\mu}R^{(10)}_{\psi\psi}+\frac{e^{-2(A-C)}}{2}\nabla^{\alpha}{\cal F}_{\alpha\mu},~~~~R^{(10)}_{i\psi}= R_{i\psi}^0+V_i R^{(10)}_{\psi\psi},~~~~R^{(10)}_{\mu i}=- {\cal A}_{\mu} R^{(10)}_{i\psi}+\frac{e^{-2(A-C)}}{2}V_i\nabla^{\alpha}{\cal F}_{\alpha\mu},\nn\\[2mm]
R^{(10)}_{\mu\nu}&= R^0_{\mu\nu}-\frac{1}{18}g^{(5)}_{\mu\nu}{\cal F}^2+ e^{-2(A-C)}{\cal A}_{(\mu} \nabla^{\alpha}{\cal F}_{\nu)\alpha}+{\cal A}_{\mu}{\cal A}_{\nu}R^{(10)}_{\psi\psi}+\frac{1}{2}\Big(\frac{1}{3}-e^{-2(A-C)}\Big){\cal F}^2_{\mu\nu},
\end{align}
where the superscript $0$ means these terms are independent of ${\cal A}$ and so they take the form they do for AdS$_5$ vacua, we have made use of the equations of motion of $d=5$ supergravity so we have $g^{(5)}_{\mu\nu}$ rather than $R^{(5)}_{\mu\nu}$ appear and so on, specifically
\begin{align}
R^0_{\psi\psi}&=\frac{e^{4C}}{2}(dV)^2 -e^{-(5A+C)}\nabla_i(e^{5A}\nabla^i(e^C)),\nn\\[2mm]
R^0_{i\psi}&=-\frac{e^{-(5A+C)}}{2}\nabla^k(e^{5A+3C}dV)_{ki},\nn\\[2mm]
R^0_{\mu\nu}&=-\big(4m^2+\frac{1}{5}e^{-3A-C}\nabla_i(e^{C}\nabla^i e^{5A})\big)g^{(5)}_{\mu\nu},\nn\\[2mm]
R^0_{ij}&= R_{ij}-(5\nabla_i\nabla_j A+ \nabla_iA\nabla_j A+\nabla_i\nabla_j C+ \nabla_iA\nabla_j C)-\frac{e^{2C}}{2} dV^2_{ij}.
\end{align}
One can also show that the dilaton terms decompose as
\begin{align}
e^{\Phi}(\nabla^{(10)})^2(e^{-\Phi})&=e^{\Phi-(5A+C)}\nabla_i(e^{5A+C}\nabla^i(e^{-\Phi})),~~~\nabla^{(10)}_{\psi}\nabla^{(10)}_{\psi}\Phi=e^{2C}\nabla_i \Phi\nabla^i C,\nn\\[2mm]
\nabla^{(10)}_i\nabla^{(10)}_j\Phi&=- V_i V_j \nabla^{(10)}_{\psi}\nabla^{(10)}_{\psi}\Phi+2 V_{(i}\nabla^{(10)}_{j)}\nabla^{(10)}_\psi\Phi+\nabla_i\nabla_j\Phi,~~~\nabla_{\mu}^{(10)}\nabla_{\psi}^{(10)}\Phi=-{\cal A}_{\mu}\nabla^{(10)}_{\psi}\nabla^{(10)}_{\psi}\Phi\nn\\[2mm]
\nabla^{(10)}_{i}\nabla^{(10)}_{\psi}\Phi&=\frac{1}{2}e^{2C}\nabla^k\Phi(dV)_{k i}+\nabla^{(10)}_{\psi}\nabla^{(10)}_{\psi}\Phi,~~~\nabla^{(10)}_{\mu}\nabla^{(10)}_{i}\Phi= -{\cal A}_{\mu} \nabla^{(10)}_i \nabla^{(10)}_{\psi}\Phi,\nn\\[2mm]
\nabla^{(10)}_{\mu}\nabla^{(10)}_{\nu}\Phi&=e^{2A} g^{(5)}_{\mu\nu}\nabla^i A\nabla_i\Phi+ {\cal A}_{\mu}{\cal A}_{\nu}\nabla^{(10)}_{\psi}\nabla^{(10)}_{\psi}\Phi
\end{align}
To show that Einstein's equations are implied we need several identities, we shall quote them as they become relevant but they can all be derived from the bi-linears in section \ref{sec:onAdS5vaccua}. First we consider ${\cal E}_{\psi\psi}$, for this one can show that
\beq
(\iota_{\xi}\psi^1_{\pm})^2= \frac{1}{16}||\xi||^2 (e^{2A}c\pm ||\xi||^2)~~~~\Rightarrow~~~~ 
\frac{e^{-\Phi}}{4}(F)^2_{\psi\psi}=  \frac{e^{-\Phi}}{4}(F)^2_{\psi\psi}\bigg\lvert_{{\cal A}=0}+  \frac{e^{-4A}||\xi||^2}{162 c^4}{\cal F}^2,
\eeq
 It then follows that
\beq
{\cal E}_{\psi\psi}= {\cal E}^0_{\psi\psi}+\frac{e^{-4A}}{162}\bigg[e^{4C}-\left(\frac{||\xi||}{3c}\right)^4\bigg]{\cal F}^2,
\eeq
the first term vanishing because it is precisely as it is for AdS$_5$ and the second because $3ce^{C}=||\xi||$ for our background. Next we consider ${\cal E}_{\mu\nu}$, here we have
\beq
H_{\mu}=-{\cal A}_{\mu} H_2-\frac{1}{3c}\tilde{\xi}\wedge {\cal F}_{\mu},~~~~ F_{\mu}=-{\cal A}_{\mu} F_{\psi}-\frac{4}{3c} e^{-\Phi}\bigg[{\cal F}_{\mu}\wedge \text{Im}\psi^1_+-e^{A}(\star_5 {\cal F})_{\mu}\wedge \text{Im}\psi^1_-\bigg],
\eeq
where one needs to bear in mind that, in addition to itself, $(\star_5 {\cal F})_{\mu}$ can contract with the ${\cal F}$ term within $F_{\psi}$. Through the identities
\begin{align}
&\sum_{k}\frac{1}{k!}(\iota_{\xi}\text{Im}\psi^1_+)_{a_1...a_k}(\text{Im}\psi^1_-)^{a_1...a_k}=\frac{1}{8}||\xi||^2 e^{A}c,~~~~(\text{Im}\psi^1_+)^2=(\text{Im}\psi^1_-)^2=\frac{e^{2A}c^2}{8},\nn\\[2mm]
(\star_5 {\cal F})^2_{\mu\nu}&={\cal F}^2_{\mu\nu}- g_{\mu\nu}{\cal F}^2,~~~~\frac{1}{2}{\cal A}_{\mu}{\cal F}_{\alpha\beta}\star_5{\cal F}_{\nu}^{~\alpha\beta}=\frac{1}{4}{\cal A}_{\mu}\epsilon_{\nu\alpha_1...\alpha_4}{\cal F}^{\alpha_1\alpha_2}{\cal F}^{\alpha_3\alpha_4}\label{eq:exidenities}
\end{align}
one establishes that
\begin{align}
H_{\mu\nu}^2&=  {\cal A}_{\mu}{\cal A}_{\nu} (H)^2_{\psi\psi}+\frac{e^{-2A}||\tilde{\xi}||^2}{9c^2}{\cal F}^2_{\mu\nu},\\[2mm]
e^{\Phi}(F)^2_{\mu\nu}&=e^{\Phi}(F)^2_{\mu\nu}\bigg\lvert_{{\cal A}=0}+{\cal A}_{\mu}{\cal A}_{\nu}e^{\Phi}( F)^2_{\psi\psi}-\frac{e^{-2A}||\xi||^2}{27c^2}{\cal A}_{(\mu}\epsilon_{\nu)\alpha_1...\alpha_4}{\cal F}^{\alpha_1\alpha_2}{\cal F}^{\alpha_3\alpha_4}+\frac{2}{9}(2 {\cal F}^2_{\mu\nu}-g^{(5)}_{\mu\nu}{\cal F}^2)\nn
\end{align}
and so after substituting for $e^C$ we find
\beq
{\cal E}_{\mu\nu}={\cal E}^0_{\mu\nu}+{\cal A}_{\mu}{\cal A}_{\nu}{\cal E}_{\psi\psi}+\frac{e^{-2A}||\xi||^2}{9c^2}{\cal A}_{(\mu}{\delta\cal F}_{\nu)}+\frac{1}{18}\bigg[1-\frac{1}{e^{2A}c^2} (||\xi||^2+||\tilde\xi||^2)\bigg]{\cal F}^2_{\mu\nu},
\eeq
where $||\xi||^2+||\tilde\xi||^2= e^{2A}c^2$ holds in general so the term in square brackets is zero, we already established ${\cal E}_{\psi\psi}=0$ and ${\cal E}^0_{\mu\nu}$ takes exactly the same form as it does for AdS$_5$ so also vanishes. The final term is simply the equation of motion of ${\cal F}$ written in component form, i.e.\
\beq
{\delta\cal F}_{\nu}= \nabla^{\mu}{\cal F}_{\mu\nu}+\frac{1}{12}\epsilon_{\nu\alpha_1...\alpha_4}{\cal F}^{\alpha_1\alpha_2}{\cal F}^{\alpha_3\alpha_4},
\eeq
clearly this should vanish  by definition.  At this point we have provided a detailed  proof of sufficiency for a solution when external supersymmetry holds, this follows from the integrability  arguments of \cite{Legramandi:2018qkr,Giusto:2013rxa}.

For a detailed proof of the fact that any solution of the external theory gives rise to a solution in massive Type IIA supergravity we must show that the remaining components of Einstein's equations and the dilaton equation of motion are implied.  The first thing we need to consider is ${\cal E}_{i\psi}$ as this appears in later expressions. For this and the other $i$ dependent terms it is useful to define the following vielbein on M$_4$ 
\beq
\e^{\underline{i}}= \Big(\text{Re}z,\text{Im}z, \e^{\perp},\frac{1}{||\tilde\xi||}\tilde{\xi}\Big)^{\underline{i}}=\Big(\text{Re}z,\text{Im}z, \frac{1}{|a|}\text{Im}(a w),\frac{b}{|a|}\text{Re}(a w)+|a| V\Big)^{\underline{i}},
\eeq
then for a general $k$-form we can decompose 
\beq
(C_k)_i=(C_k)_{\psi}V_i+ \e^{\underline{i}}_i(\iota_{\e^{\underline{i}}}C_k).
\eeq
The important identity for the component at hand is
\beq
\sum_{k}\frac{1}{k!}\big((\text{Im}\psi^1_{\pm})_{\psi}\big)_{a_1...a_k}(\iota_{\e^{\underline{i}}}\text{Im}\psi^1_{\pm})^{a_1...a_k}=0,\label{eq:forEitau}
\eeq
it is then simple to establish that 
\beq
{\cal E}_{i \psi}= {\cal E}^0_{i \psi}+V_i {\cal E}_{\psi\psi}\, ,
\eeq
with each term in the sum necessarily zero. Next for ${\cal E}_{\mu\psi}$ we need only reuse the identities in \eqref{eq:exidenities}, we find
\beq
{\cal E}_{\mu\psi}={\cal E}^0_{\mu\psi}- {\cal A}_{\mu}{\cal E}_{\psi\psi}+ \frac{e^{-2A}||\xi||^2}{18c^2} \delta {\cal F}_{\mu},
\eeq
with all terms in the sum again zero, at least when the external flux equations of motion is assumed to hold.
For ${\cal E}_{\mu i}$ we can make use of \eqref{eq:forEitau} again and the identities
\begin{align}
\sum_{k}\frac{1}{k!}(\iota_{\e^{\underline{i}}}\text{Im}\psi^1_{+})_{a_1...a_k}(\text{Im}\psi^1_{-})^{a_1...a_k}=0,~~~~\sum_{k}\frac{1}{k!}(\iota_{\e^{\underline{i}}}\text{Im}\psi^1_{-})_{a_1...a_k}(\text{Im}\psi^1_{+})^{a_1...a_k}=\frac{e^A c}{8}||\tilde{\xi}||\delta_{\underline{i}\underline{4}}
\end{align}
from which it follows that
\beq
{\cal E}_{\mu i}=-{\cal A}_{\mu} {\cal E}_{i \psi}+\frac{e^{-2A}||\xi||^2}{18c^2} V_{i} \delta {\cal F}_{\mu}
\eeq
which is implied by the equation of motion of ${\cal F}$ and the previous conditions. The final component of Einstein's equations is ${\cal E}_{ij}$, we have already quoted all but one of the required identities to tackle this, namely
\beq
\sum_{k}\frac{1}{k!}(\iota_{\e^{\underline{i}}}\text{Im}\psi^1_{\pm})_{a_1...a_k}(\iota_{\e^{\underline{j}}}\text{Im}\psi^1_{\pm})^{a_1...a_k}=\frac{1}{16}(e^{2A}c^2\delta_{\underline{i}\underline{j}}\mp||\tilde{\xi}||^2\delta_{\underline{i}\underline{4}}\delta_{\underline{j}\underline{4}}),
\eeq
we find this decomposes as
\beq
{\cal E}_{ij}={\cal E}^0_{ij}-V_iV_j {\cal E}_{\psi\psi}+2 V_{(i}{\cal E}_{j)\psi},
\eeq
with each term in the sum again zero --- this exhausts all components of Einstein's equations.  Fortuitously establishing that the dilaton equation of motion is implied is a much shorter computation, we find
\beq
{\cal D}= {\cal D}^0+\frac{e^{-4A}}{9 c^2}(e^{2A}c^2-\xi^2){\cal F}^2- e^{-4A}\frac{1}{9 c^2}\tilde{\xi}^2{\cal F}^2
\eeq
which is solved due to $||\xi||^2+||\tilde\xi||^2= e^{2A}c^2$ and the fact that ${\cal D}^0=0$ because ${\cal D}^0$ is precisely equal to the  ${\cal D}$ of the AdS$_5$ vacua. Thus, the embedding of any solution of $d=5$ minimal gauged supergravity, not merely the supersymmetric ones, into massive Type IIA supergravity always yields a solution of the Type IIA supergravity equations of motion.

\bibliographystyle{JHEP}

\bibliography{IIAreductionbib}

\end{document}